**Structural matters in HTSC; the origin and form of stripe organization and checker boarding**

John A. Wilson

H.H. Wills Physics Laboratory,

University of Bristol,

Bristol BS8 1TL

United Kingdom

**Abstract**

The paper deals with the controversial charge and spin self-organization phenomena in the HTSC cuprates, of which neutron, X-ray, STM and ARPES experiments give complementary, sometimes apparently contradictory glimpses. The examination has been set in the context of the boson-fermion, negative-$U$ understanding of HTSC advocated over many years by the author.

Stripe models are developed which are $2q$ in nature and diagonal in form. For such a geometry to be compatible with the data rests upon both the spin and charge arrays being face-centred. Various special doping concentrations are closely looked at, in particular $p = 0.1836$ or $^9/_{49}$, which is associated with the maximization of the superconducting condensation energy and the termination of the pseudogap regime. The stripe models are dictated by real space organization of the holes, whereas the dispersionless checkerboarding is interpreted in terms of correlation driven collapse of normal Fermi surface behaviour and response functions. The incommensurate spin diffraction below the 'resonance energy' is seen as in no way expressing spin-wave physics or Fermi surface nesting, but is driven by charge and strain (Jahn-Teller) considerations, and it stands virtually without dispersion. The apparent dispersion comes from the downward dispersion of the resonance peak, and the growth of a further incoherent commensurate peak ensuing from the falling level of charge stripe organization under excitation.

## §1. Introduction

I have for a long time been trying to get the HTSC community to focus on a mixed-valent, two-subsystem, negative-$U$ approach to cuprate superconductivity [1]. This has entailed going through the entire panoply of highly relevant data, whether the NMR and μSR data, or the laser pump-probe and thermomodulation optical data, or the transport data such as the recent Nernst work, or again the specific heat data. From the beginning I have advocated that the systems are inhomogeneous, and have continually striven to show what it is that makes mixed-valent cuprates unique within the Periodic Table, besides inserting further matters of 'chemistry' such as disproportionation, the Jahn-Teller effect, and the role of the ionicity of the various counter-ions. The present work proceeds along the same lines and re-addresses the details of charge stripe behaviour, now treated on a diagonal 2$q$ basis. Both the hole-charge array and the attendant discommensurate antiferromagnetic spin array each emerge as being face-centred, which accounts for the diffraction generated by these incipient orderings and detected in neutron and synchrotron X-ray work. Sections 2 and 3 introduce and develop some of the topics necessary for addressing this matter appropriately, such as inhomogeneity, ionicity and p/d mixing, the spin and charge gapping and pseudogapping, structural adjustments and lattice coupling. The important role of synchrotron work in directly revealing charge ordering is stressed, and the results are connected through to those obtained for the nickelates. The difference between 1-$q$ and 2-$q$ incommensurate behaviour is stressed. Section 4 examines the ordering encountered at selected key doping levels, and in particular compares the circumstances at $p = 0.157$, where $T_c$ maximizes, with $p = 0.184$, where the superconducting condensation energy sharply maximizes. What is occurring at these various compositions is related through to the boson-fermion negative-$U$ modelling advanced over many years by the author. In section 5a the new checkerboarding phenomenon is addressed and is interpreted in terms of correlation-driven Fermi surface boxing and unzippering in the region of the extended saddles. Correspondingly in section 5b it is asserted that the reported IC magnetic behaviour does not in the HTSC cuprates result from standard Fermi surface nesting-dictated physics. The strong input of the axial saddles close to $E_F$ is apparent throughout, and the latter clearly act to the overall benefit of HTSC rather than to its detriment.

## §2. The electronic setting of local pair superconductivity in the cuprates.

### 2.1) Concerning structural responses to electronic counts, shell closure and negative-$U$ states.

I would like to begin the present paper by considering that simplest of matters, namely the lattice parameters, and what these can disclose about the HTSC condition. Recently Röhler has assembled [2] an extensive collection of high resolution crystallographic data taken from a diverse set of HTSC systems and has presented them in a form which offers remarkably direct insight into what is afoot in these materials. Rather than plotting the lattice parameters *a*, *b* and *c* individually as a function of composition, Röhler has sought to avoid the various complications associated with the Jahn-Teller effect, octahedral tilting and *c*-axis charge reservoir effects by concentrating on the basal area of the planar $CuO_2$ unit, $B(p)$. In underdoped systems stress directed along the basal plane has the remarkable capacity to force $T_c$ up strongly: indeed for YBCO-124, $T_c$ rises from 80 K to 105 K under a uniaxial pressure of 10 GPa, and this with relatively little change actually being incurred in the basal parameters [3]. The very striking observation to be gained from Röhler's presentation of the crystallographic data is that across the central superconducting range ($p \sim 0.10$ to $0.22$), $B(p)$ always is in clear excess of what a simple extrapolation in from the two outer wings of the data plots would yield. This *upward* deviation of the data within the central region strikingly mimics the well known rise and fall of $T_c$, $B(p)$ similarly peaking near $p \sim 0.16$. What is more, between the various different HTSC systems the magnitudes of the $B(p)$ humps reflect the differing magnitudes of $T_c^{opt}$. Note the above correlations are abstracted from *room* temperature data.

The decrease of lattice parameter occurring from $La_2ZnO_4$ to $La_2CuO_4$ to $La_2NiO_4$ results from the progressive fall in occupancy of the *anti*bonding σ* Cu(3*d*)-O(2*p*) $e_g$ band with $d_{x^2-y^2}$ symmetry from 2 to 1 to 0. These three occupancies are associated respectively with the formal divalent cation configurations $d^{10}$, $d^9$, and $d^8$, or, using my own preferred expanded notation introduced in [1a,b], $^{10}Zn_{II}^0$, $^{9}Cu_{II}^0$, and $^{8}Ni_{II}^0$. In the case that such a reduction in the number of σ* electrons within the system actually is procured through an increase in the average valence of the T.M. ion, as with the '*hole*-doped' mixed-valent cuprates, the ensuing decrease in the lattice parameters becomes even more pronounced, due to the covalent action of the bonding term being now augmented by an ionic one. Thus in LSCO $B(p)$ drops from 14.36 Å$^2$ at $p = 0.07$ to 14.18 Å$^2$ at $p = 0.30$, a drop of 1¼%. It is this to be expected decline in $B(p)$ which now is demonstrated in the system holding the highest $T_c$ dealt with by Röhler [2] – namely Hg-1212 – actually becoming completely negated over the central key range $p = 0.12$ to $0.21$, prior to staging a precipitous recovery in the anticipated behaviour beyond $p \approx 0.24$.

Our earlier work encourages one here to view the hump found in $B(p)$ as the product of a local pair bosonic population, the latter acquired in fluctuational style within a system displaying near-resonance between the quasiparticle Fermi energy and the local-pair ground-state [1]. The pair involved here is, note, a two-electron one, not a two-hole one as is so often presumed by others [4]. In our previous treatment [1b] the above key bosonic fluctuation is designated $^{10}Cu_{III}^{2-}$. This transient two-electron cell loading becomes sensed in the crystallographic data by reason of the locally generated $p^6d^{10}$ count. The shell filling brings the termination here of all strong p/d bonding within the trivalent coordination unit acquiring the two additional electrons. For the customary two-*hole* boson state one would conversely see a *de*crease in lattice parameter (*i.e.* see a dip in $B(p)$) as one gains a local condition in which the individual polaronic hole carriers (site hybridized $^8Cu_{III}^0/^8Cu_{II}^{1+}$) become bound together into pairs as two-site bipolarons [5].

**2.2) Concerning highly local, inhomogeneous electronic behaviour.**

That the above mixed-valent substitution in the HTSC cuprates generates significant inhomogeneous electronic behaviour at the atomic level is today surely incontestable. Improved analyses of the NMR [6] and μSR [7] data, plus the remarkably graphic energy-resolved scanning tunnelling microscope results [8], all clearly register the high degree of local disturbance and intrinsic variability experienced in these systems to beyond optimal doping. Indeed it is not really justified to contemplate using mean-field approaches and standard Fermi liquid theory until one has passed entirely beyond the superconducting range. The most basic of transport measurements serve to show the chronic level of scattering experienced by the charge carriers right across the range, and from which not even the nodal carriers are immune [1c,9]. The Seebeck coefficient, for example, makes very apparent the persisting degree of development of a density-of-states pseudogap even as $T_c$ maximizes [10]. Indeed the fact that the carriers respond at low temperatures as holes throughout the HTSC range puts in evidence how far correlation and site disorder are pushing the systems to a localized condition. This in the II/III cuprates is in spite of having $E_F$ sited within the *lower* half of what standard LDA band structure calculations would present as an effectively sinusoidal 2D band with the latter supporting an almost circular Fermi surface. ARPES results [11] provide a clear record of the progressive disintegration of Fermi liquid behaviour upon decrease in the dopant concentration as one retreats towards the properties of the undoped parent $d^9$ Mott insulator. Accordingly it would seem quite inappropriate to appeal to rather simple Fermi liquid nesting physics when attempting, for example, to interpret the magnetic behaviour of HTSC materials. The superconductivity itself, indeed, makes apparent the local character of these materials in its coherence lengths, $\xi$, of just 3 or 4 unit cells [12].

**2.3) Concerning hole concentration dependence of pair density, condensation energy and $T_c$.**

Complementary to the short coherence lengths of HTSC materials come their long penetration depths, $\lambda$. $\lambda$ may be readily extracted from the μSR measurements [13], and these provide thereby a direct monitor of the superconducting pair concentration, $n_s$ ($\propto \lambda^{-2}$). Highly significantly this pair population is dictated by the 'hole' dopant concentration $p$ ($n_s = p/2$) over quite an extensive range, and not by the complementary electron count, 1-$p$. Yet more revelatory of HTSC behaviour is that, once clear of the superconducting onset around $p = 0.05$ [14], $T_c$ is found to rise steadily with $n_s(p)$ up to about $p = 0.125$ [15]. Beyond the latter doping level, first $T_c(p)$ comes slowly to a maximum – this universally at $p = 0.16$ according to the empirical findings of Tallon *et al* [16] – shortly after to be followed near $p = 0.19$ by $n_s(p)$ as well. The strong fall back in $n_s(p)$ which follows upon still further increase in $p$ has become known as the 'boomerang' effect. This can most fully be seen in the Tl-2201 system, which is subject to high overdoping in consequence of its complex defect structure [16]. The effect does not result from a decrease in $m^*$, which would carry the μSR signal in the opposite direction. The above turning points in $T_c(p)$ and shortly thereafter in $n_s(p)$ arise naturally within the resonant B-F modelling of HTSC [1].

From the analysis of the electronic specific heat work carried through by Loram et al [17], one can witness an effect directly related to the above; namely of seeing $T_c(p)^{max}$ come slightly in advance of the concentration $p$ (~ 0.185) at which $U_c(p)$, the superconducting condensation energy per gm-atom, mounts sharply to its maximum. Here $T_c(p)$ is more a monitor of the effective pairing strength, whereas $U_c(p)$ tracks the progress of the entire condensation, not only condensed local pairs, but also those additional quasiparticles brought into Cooper pairing under the resonant action of the local pairs [1,18,19,20]. At the point where $T_c$ maximizes ($p \approx 0.16$), the HTSC cuprates still show a fair degree of pair breaking coming from the level of remnant magnetism [7,21]. This capacity for pair breaking is dropping away very rapidly however with $p$, and the most effective composition for attaining maximization of $n_s$ and $U_c$ is delayed only fractionally behind that for $T_c$. The reason why $T_c$ universally optimizes near $p = 0.16$ is that this value constitutes a particular geometrical criterion in the interrelation between the two subsystems existing within the square-planar, mixed-valent materials with regard to electronic microstructure and percolation [1i].

**2.4) Concerning boson-fermion resonance, ionicity, Jahn-Teller effect, pair breaking and spin gap.**

To achieve optimization of $T_c$ between the *different* HTSC systems calls for (i) reaching the resonance between the local pair negative-$U$ state and the chemical potential at precisely the above critical $p$ value, and (ii) having the magnetic pair breaking in the system at this stage as low as is feasible, limiting thereby the delay in the peaking of $n_s(p)$ and $U_c(p)$ vis-à-vis $T_c(p)$. For HTSC systems the way to suppress the pair-breaking capacity is to increase the general covalence. Towards this end one may pursue two different courses [1f,g]: (i) introduce counter-ions into the system of a high atomic number, like Hg, Tl, Pb, and Bi, so augmenting the overall dielectric constant and screening, and (ii) generate chemical pressure within the basal plane through promoting the Jahn-Teller distortion of the $^9Cu_{II}^{0}$ divalent subsystem. It is discovered that mercury offers the best option here. Indeed the Jahn-Teller distortions in the mercury cuprates reach quite exceptional values – for the case of Hg-1201, basal and apical Cu-O bond lengths of 1.94 Å vs. 2.80 Å, respectively. What the Jahn-Teller deformation of the Cu-O coordination unit brings about is compression of the $CuO_2$ basal plane, via added stabilization of the σ$d/p$ $d_{x2-y2}$ symmetry bonding band, this to the advantage of the full *anti*bonding $d_{z2}$ σ*$d/p$ band. The beneficial shrinkage of the crucial outermost $CuO_2$ planes (*i.e.* those proximate to the excess hole-introducing oxygen atoms) is augmented in the earlier of the 'higher order' HTSC systems by the additional compacting action of the *four*-fold coordinate, inner planes. The above proceeds up to such a stage as magnetism re-emerges within the now virtually undoped innermost layers. At optimal doping, $a_o$, the cell edge, from its 3.882 Å in Hg –1201, becomes 3.859 Å in Hg-1212, 3.852 Å in Hg-1223, and 3.850 Å in Hg-1234, while the corresponding $T_c^{max}$ values are 97 K, 127 K, 135 K and 127K, with a further fall in Hg-1245 to 110 K [22].

What the above advance in covalence specifically secures as regards pair breaking is a reduction in the effective magnetic moment formed at magnetically active sites [23] and the emergence, moreover, of spin-singlet spin-gapping via a promotion of dynamic RVB coupling [24] at the expense of frozen Néel order. The size of the spin gap developed hence is smallest in LBCO and LSCO [25] and rises through $YBa_2Cu_3O_{6+x}$ [26] to the more weakly ionic HTSC systems, like Tl-2201 [27] and Bi-2212 [28]. One way to counter the above development is, of course, to apply an external magnetic field [29,30] (a subject we return to later). Perhaps a more unexpected and revealing way to promote pair breaking is to substitute some Cu in the planes by Zn. The effect is locally to free up the spins around each substituent centre, this causing $T_c$ to drop particularly rapidly. The process has been examined through μSR [31], transport [32], neutron scattering [33], and many other measurements. The $p$ value providing the maximization of $T_c(p,y)$ becomes steadily forced up with the Zn substitution level, $y$ [33], towards where $n_s(p)$ is maximized near $p$ = 0.185, to obtain relief from the severity of pair breaking and a restoration of delocalization.

**2.5) Concerning the tight-binding bandstructures, the level of *p/d* mixing and screening.**

$YBa_2Cu_3O_{6+x}$, although so widely studied, introduces added complication by reason of the nature of its 'charge reservoir', the partially oxygenated Cu-O chains. These participate more intimately in events than is the case for most other HTSC systems. In particular the copper chain sublattice ultimately shares in the superconductivity, although the μSR results clearly reveal that the many paired carriers finally to appear in the chain regions of the structure do not significantly contribute towards advancing $T_c$ [34]. Only the hole population transferred through from the reservoir onto the planes confers the HTSC response. By $YBa_2Cu_3O_{7.0}$ the system has become just slightly overdoped ($p$ ~ 0.17) in the sense that $T_c$ has just peaked. It is possible to push the planar hole content to still higher levels by making the counter-ion substitution of $Ca^{2+}$ for $Y^{3+}$, and in this way to access the overdoped planar regime of falling $T_c$, $n_s$ and $U_c$ [35, 36,17]. Independently the chains in the system meanwhile are free to display CDW behaviour [37], this in addition to showing transverse structural ordering of their variable oxygen atom content [38].

Because $YBa_2Cu_3O_7$ itself is stoichiometric and is relatively metallic, even a standard LDA band structure calculation can offer some useful insight into the electronic circumstances prevailing within mixed-valent HTSC systems in general. Such calculations immediately reveal the high degree of tight-binding and ionic behaviour prevalent in these oxides. When the wavefunctions appertaining to the entire valence band are summed, the bulk of those electrons are indicated to reside in quasi-ionic, spherical distributions [39]. If only the particular wavefunctions present at $E_F$ are summed, one can see furthermore how those electrons become distributed in real space. The two much fuller $d_{x2-y2}$ bands of the three (Z = 1) site their electrons primarily within the planes, whilst the third and much *emptier* $d_{x2-y2}$ band disposes of its electrons almost exclusively on the chains. This means the square-planar coordination units display the lower formal valence. One observes, however, that there is clear register within these plots of the emergence of one sixth of a *hole* within each of the two planar Cu-O units per formula unit. The complementary one third of an electron being transferred through onto the chains per Cu there then sources the metallicity, the CDWs and the superconductivity appertaining to that segment of the structure [1f].

What such band structures and their associated charge plots make very evident for the cuprates is the particularly high degree of *p/d* mixing current in HTSC oxides. This mixing is due to the imminent closure of the copper *d*-band and the already strongly descending (*i.e.* stabilizing) energy of all the *d*-subbands [1i,40]. Within the *p/d* hybridized σ/σ* manifold of states with $d_{x^2-y^2}$ symmetry, considerable oxygen *p* weighting has become inverted from the fully occupied σ band into the partially occupied σ* band. Even for a purely *divalent* cuprate, this *p/d* admixing already is such that the upper and here partially occupied wavefunction of the σ* band bearing $E_F$ runs extensively over the oxygen as well as the copper sublattice; *i.e.* even at this ($^9Cu_{II}^0$) loading the *p*-shell is not exhibiting the closed $p^6$ condition. Note this non-closure is not the result of having $E_F$ enter the top of the non-bonding $p_z$ band. Turning to a mixed-valent system undergoing augmentation of its average valence, the admixed *d*-states now bear proportionately even less of the introduced hole than do its already opened $p_{x,y}$-states. The quasiparticles in the cuprate II/III systems indeed run quite as freely over the oxygen sites as they do over the copper sites in the cell. One encounters accordingly far greater metallicity in such cuprates than is the case with the analogous nickelates and manganates. The great benefit of this freeing up of the spatial extent available to the relevant wavefunctions is that there results appreciable reduction in the coulombic repulsion between carriers from the more effective screening.

The above advanced level of *p/d* mixing in the cuprates, and how it adjusts to changes in the hole doping concentration, very recently has been directly evaluated by NMR, both for LSCO and for YBCO [41]. These measurements register very clearly how the fresh space opened up to conduction actually runs largely over the oxygen sublattice. This fact re-emphasizes why it is so important when designating the hole doping centres to employ $Cu_{III}$ as preferred book-keeping device for the number of electrons in play, it referring to the entire coordination unit, and to avoid as far as possible the frequently used symbols $Cu^{3+}$ or $d^8$. All copper chemistry tells one that it is never possible to access anything like $d^8Cu^{3+}$ outside a hard Mott insulator like $KCuF_4$ (see fig.3 in [1i]). This essential chemistry, coming from the imminence of shell closure, is missed in much simple numerical modelling of the cuprates. The key outcomes to follow in the HTSC materials are (i) the facilitating of $p^6d^{10}$ two-electron shell-closure fluctuations and transient negative-*U* pair formation, (ii) the constraining of spin fluctuations and their pair-breaking activity.

**2.6) Concerning disproportionation and less extreme structural adjustments.**

The above situation relating to *p/d* mixing when extended to second and third TM series group VIIIc materials procures full-blown static disproportionation in divalent AgO [42] and AuO or $CsAuCl_3$ [43]. By contrast for the first series *mixed-valent* II/III cuprate systems, the potential self-organization states are the somewhat less extreme ones of electronic stripe formation, of clustering and, possibly, of its converse, Wigner crystallization. The latter options disturb less the underlying structure of the lattice. As was noted earlier, a good deal of structural activity is, nonetheless, current right up to optimal doping, as is registered under the local electronic probes of NMR [6], µSR [7], and energy-resolved STM [8], or again in the direct lattice probings afforded by the EXAFS [44], PDF (pair distribution function) [45] and D-W (Debye-Waller) [46] techniques. What are in question now are the time scales and the degrees of self-ordering actually attained.

All HTSC systems display some tendency towards such electronic ordering around $p = \frac{1}{8}$ [47,1e], and this in fact is statically acquired in LBCO [48,49], with possible assistance from a slight structural adjustment in the tilting of the Cu-O octahedra. The latter structural change (from LTO to something akin to LTT) precedes by about 10K the onset in LSCO of the detected electronic ordering. The latter ordering clearly has both magnetic and charge components to its character. Since all HTSC systems would appear to exhibit some susceptibility to a $\frac{1}{8}$th event, and to what early neutron diffraction work indicated to involve stripe segregated charge and magnetic phase slip, we next shall examine more closely just what current diffraction work, neutron and X-ray, indicates that incipient order might be. Any geometrical micro-ordering clearly is of much potential import towards understanding the details of HTSC. As has been observed already, the point where $T_c$ universally maximizes is set by the geometric criterion of percolation, whilst the above $\frac{1}{8}$ anomaly would look in LBCO and LSCO to pick up commensuration energy associated with the "LTT" structuring. It is necessary finally to recall that in high quality LBCO, when the $\frac{1}{8}$ anomaly structure actually becomes frozen in, it becomes associated with a total suppression of HTSC [50].

### §3. Experimental input into resolving 'stripe phase' structuring.

**3.1) Concerning charge stripe orientation and lattice coupling through the Jahn-Teller effect.**

Because in LBCO the $\frac{1}{8}$ structure actually freezes in, it is possible to investigate the frozen structure quite closely from existing neutron and X-ray data. Immediately it is evident from the relative elastic neutron scattering line-widths in [49] that the degree of magnetic organization developed ultimately is of a higher level than is the charge order. That matches the fact that the $^9Cu_{II}^0$ sites carrying the magnetic moments form the majority species. The moment-free minority $^8Cu_{III}^0$ species being more sparsely distributed are less strongly coupled. It has become commonplace to speak of such organization in terms of stripes as distinct from clustering, and indeed, as the materials customarily are quenched from high growth temperatures, there is little evidence of any clustering of the substitutional counter-ions themselves. The maintenance of local charge neutrality indeed is against this. What drives the observed ordering in the mixed-valent cuprates, as with the corresponding manganates and nickelates, appears to be the strain field associated with the strong Jahn-Teller activity in the σ/σ* bands of $e_g$-related symmetry. Accordingly the ordering reflects more the adopted locations of the dopant electrons than it does the exact locations of the dopant ions. Within the LSCO and Hg-1201 structures the dopant ions are directly proximate to four Cu atoms in a checkerboard plane. As examination of fig. 4 in [1i] will indicate, the electrons then have to do little more than to settle preferentially into one or other of these four coordination units in order to cause substantial organization along the above lines.

A quasi-linear response becomes the natural way to effect self-organization of the excess charge and at the same time bound the magnetic moment bearing domains. Such organization is well suited to canted runs of J-T distorted $^9Cu_{II}^0$ octahedra, these capable of accommodating to the more regularly shaped octahedra associated with the hole-bearing stripes which remain untilted [51]. In the orthorhombic LTO structure of the pure divalent parent, $La_2CuO_4$ itself, canting already occurs, with the canting axis running in a particular 45° direction (neglecting twinning). What happens in the much examined case of $p = \frac{1}{8}$ LBCO is that at the crystallographic transition near 60K the unit cells revert from being uniformly orthorhombic to being on average tetragonal. This change introduces an averaged basic cell now of precisely √2 x √2 ($a_o$ HTT), as the octahedral tilting of the LTO state fragments from occurring in uniform rows parallel to a *single* 45° axis (per orthorhombic domain) – first clockwise, then anticlockwise – into following a more complex pattern of local site tilts about both diagonal axes. This passage from a single- to a two-axis condition is the structural forerunner of the charge and spin organization duly to emerge within the *electronic* system of the mixed-valent material. The above pattern of LTT tilts is such that the associated unit cells retain 45°-rotated form, and one might well anticipate that as the electronic stripes develop they follow these same, now fully equivalent, orientations.

**3.2) 2-q versus 1-q behaviour and its influence on diffraction from discommensurations and stripes.**

Much as the double-axis tilts in the above l.t. structure do not betoken a 2×2, *x,y* cell, so one must be careful, when dealing with 2-*q* as distinct from 1-*q* incommensurate behaviour, not to rush to ascribe the predominantly *x/y* subsidiary spotting present in the spin- and charge-organized diffraction patterns to features running in the *x*- and *y*-axis directions. When drawing up the figures in the Appendix of [1f], or again in [1e], I myself was guilty of this mistake, and ought not to have been since we had had long experience of stripe phase structures in more than one dimension. That arose in connection with the discommensuration arrays we had discovered in the incommensurate CDW phases of $2H-TaSe_2$, *etc.*. Figure 10 in [52a] or figure 2 in [52b] serve to illustrate how, for hexagonal geometry, the discommensurations (DCs) align themselves at 30° to the incommensurate wavevectors of the diffraction pattern. This occurs when the DCs represent simultaneous phase slips along more than one principal direction. In the square-planar case there is involved π phase slip in the $2a_o$ AF array along both the *x* and *y* axes, and it leads to overall magnetic pattern displacements of $\sqrt{2}a_o$ in the 45° directions. The DC arrays that arise now in the symmetric 2-*q* tetragonal case fortunately are considerably simpler than for the symmetric 3-*q* hexagonal case, where the three active IC vectors lead to a 'double honeycomb' domain structure [52].

In the cuprate case so far there has been almost exclusive focus on 1-q striping, stemming from Yamada *et al*'s finding that the size of $\mathbf{q}_I{'}$ is linearly proportional to *p* up to about *p* = $\frac{1}{8}$ [53,1e]. Here the $\mathbf{q}_I{'}$ are the four equivalent subsidiary wavevectors governing the incommensurateness of the magnetic structuring away from the commensurate reference vector of the parent antiferromagnetic condition. In the $p = 1/8$ case, $\mathbf{q_I}'$ becomes $(1/4, 0)$, etc. if it is expressed in terms of the vector $(\pi,\pi)$ [or $(½,½)2\pi/a_o$] for the Néel state of LCO ($a_o$ being for the HTT structure). The relatively high intensity of these magnetic satellites is subsumed from the now defunct parent AF spotting. This magnetic scattering, elastic for LBCO $p = 1/8$, is accompanied by its charge counterpart (much weaker as regards neutron scattering) with wavevectors $\mathbf{q_I}''$ of $(1/4, 0)$, etc., where the latter IC spots have this time been quoted in relation to a basic lattice vector: *i.e.* in real space they are enumerated in units of $a_o$ and not the $2a_o$ relevant to the antiferromagnetic spotting. Accordingly, here at $p = 1/8$ the *axial* charge modulation period amounts to $4a_o$, whilst the accompanying *axial* spin modulation period is twice that at $8a_o$. The charge spotting, as sensed by neutrons or X-rays, is of extremely low intensity and it derives independently from a superlatticing diffraction from the holes themselves. $1/8$ $a_o*$ diffraction spotting from the overall charge array, as we shall appreciate later, is forbidden by virtue of the axially face-centred form to the charge ordering (see §4(i)).

**3.3) Sources of electronic organization. Case of $(La_{2-x}Sr_x)NiO_4$. Magnetic order and pseudogapping.**

I hesitate to call the above periodic entities CDWs and SDWs, since I have always tried to limit those terms to states with wavelengths dictated by the Fermi surface topology, as in the transition metal dichalcogenides [54]. Personally I do not see this as being the case for the HTSC cuprates. As the ARPES results highlight [11], the Fermi surface is not well established in the latter materials, in particular around the saddles at low $p$. Furthermore, if nesting were to be occurring across individual arms of the FS (as in the 1T dichalcogenides), the CDW wave*length* would grow with $p$, not the converse observed experimentally [53]. Again if the nesting were between pairs of saddle points, as may well be the case with $2H-TaSe_2$, *etc.*, the nesting vectors should be much closer to being commensurate than are the observed $\mathbf{q_I}$. Finally if the nesting were across the body of the zone, as many would claim, the spots then would define directly the principal vectors (as say in chromium), and not the subsidiary ones – the $\mathbf{q_I}'$, associated above with discommensurate striping. We, in such a case, would be relating the spotting to $x$, the full electronic count, rather than it pointedly expressing the hole count $p$ away from half filling. Any relatively standard RPA treatment of the susceptibility seems misplaced here, especially in the light of Yamada *et al*'s telling observation, $|\mathbf{q_I}'| = p$ [53]. As I emphasized earlier in [1e,f] the latter relation speaks of some simple numerology within a real space dictated organization. The latter type of direct interaction finds excellent illustration in the charge striping arrays recently to have been recorded in detail by Ghazi *et al* [55] through synchrotron X-ray work on the analogous $(La_{2-x}Sr_x)NiO_4$ system. With the latter, and for $x$ and $q_I$ between $1/4$ and $1/3$, cooling to helium temperatures uncovers strong lock-in effects in the $\mathbf{q_I}$ to values of 0.294 and 0.313 $\mathbf{a_o}*$, besides to 0.333 $\mathbf{a_o}*$ itself (see figure 10). It is to be noted that what these lock-in values amount to are $5/17$, $5/16$ and $5/15$ precisely. They arise here under a discrete, 5-bit, row sequencing in relation to segregated hole siting, involving whether holes are confined to each third atom row or become delayed until the fourth, leading to 17, 16 and 15 row sequences of

**3** 4 3 3 4 **3** ,    **4** 3 3 3 3 **4** ,   and, of course,   **3** 3 3 3 3 **3** ,   respectively.

We shall later encounter also $5/18$ or 0.278, this associated with **4** 3 4 4 3 **4** . The above form of faulting has long been known for defect compounds and alloys, *e.g.* $Cu_{2-x}Te$ and $Ni_{3-x}Te_2$ [56], and it usually has nothing to do with the Fermiology: indeed it may occur in non-metals.

The level of local disturbance which the onset of charge stripe order itself generates amongst cells (dependent upon their individual proximity to the stripes) has numerically been assessed in a wide range of LSCO samples through an extensive PDF analysis undertaken by Billinge *et al* of their low temperature neutron scattering data [57]. Unfortunately in their modelling they presume uniaxial striping, but this should not invalidate their main finding that the *spread* incurred in the distribution of local Cu-O basal bond lengths actually is greatest for *x* just slightly above 0.15. There this spread has become 0.02 Å larger than that present in the bounding, stripe-free systems $x \approx 0$ and 0.25. The above peaking in local disturbance to the structure [fig.3a, 57], unlike the peaking discussed in §2 to the average lattice parameters over the same range of *x*, arises now only once into the low temperature charge-ordering regime. It is the direct consequence here of the stripe ordering and does not issue simply from the disorganized mixed valence alone.

With neutron work the degree to which stripe organization actually becomes apparent for a given HTSC system depends greatly on the level there to which the magnetism remains unquenched [see §2]. Magnetic discommensurations are going to be readily discernible only where RVB coupling has not been able to open up a spin pseudogap. In $YBCO_7$ that limits the condition for observing the $\mathbf{q_I}'$ striping spots to inelastic excitations of above ~ 33 meV [26] ($\equiv$ 400 K and locally equating to $T_N$ in $YBCO_6$). With LSCO in the vicinity of $p = 1/8$, the spin gap takes the much lower value of about 7 meV [25], and indeed for LBCO it has become reduced to near zero, as the superstructure actually freezes in [49]. As noted previously, the spin gap may be filled in (rather than pushed uniformly down) by the application of a magnetic field [29,30] - this inside a type II superconductor below $T_c$ quantized into flux vortices. Recent neutron work from Gilardi *et al* [58] on OD *x* = 0.17 LSCO has shown that, under an external magnetic field of just 5 tesla, a sizeable residue of magnetic scattering, strong at 40 K just above $T_c$, continues right down to 5 K even at excitation energies as low as 4 meV. In this work, $\mathbf{q_I}'$ itself was found to have increased from the $1/8 \mathbf{a_o}^*$ at $p = 0.125$ to $1/7 \mathbf{a_o}^*$, a value reported earlier in [25] using a slightly overdoped *x* = 0.165 LSCO sample.

**3.4) Using X-ray work to probe charge organization.**

Whilst most of the attention hitherto has been on neutron diffraction and the spin array, largely reflecting and indeed fostering a preoccupation with the spin fluctuation modelling of HTSC, we now shall give particular attention to the charge array information deriving from X-ray scattering. Because there we are sensing only the hole charge ordering, the resulting scattering intensity becomes extraordinarily weak and this necessitates synchrotron work. Niemöller *et al* [59] have undertaken such work at DESY, and they published their findings in 1999 in what has proved a surprisingly little quoted paper. Their work on a *x* = 0.15 $(La,Nd,Sr)_2CuO_4$ sample contains however at least a dozen points of key importance towards resolving what is going on:

1) the $q_I''$ charge spot intensity is only ~ $10^{-8}$ that of the basic Bragg spots.
2) the size of this wavevector at $x = 0.15$ is connected still with an $8a_o$ periodicity.
3) the modulation coherence length, at 44 Å, stays similarly unchanged from that for $x = 0.125$. Note the distance amounts to $\sqrt{2} \times 31$ Å, where 31 Å equals $8a_o$.
4) the peak intensities, however, are 30% weaker at $x = 0.15$ than they were at $x = 0.125$.
5) the above charge order coherence length proves to be temperature independent, and in this is unlike the magnetic coherence length which decreases continuously with temperature increase.
6) the amplitude of the charge order diffraction spots diminishes, meanwhile, approximately linearly with $T$.
7) the charge order spotting for the $x = 0.15$ sample on cooling only onsets, unlike in the $x = 0.125$ case, some 8 K below the LTO → LTT transition (now 62 K). The latter structural transition is monitored here by growth in the (3,0,0) Bragg spot, forbidden for the LTO structure.
8) once away from $x = 0.125$, the charge ordering sets in at a somewhat *higher* temperature than does the spin ordering. This very significant observation was made earlier too by Tranquada *et al.* in [48].
9) the charge order, moreover, remains in evidence down into the magnetic pseudogapping regime, where the magnetic spotting has faded away.
10) for the $x = 0.125$ case, the charge and magnetic orderings appear simultaneously triggered by the LTO → LTT structural transition.
11) below $x = 0.125$ charge ordering persists through up into the LTO condition, and far above $T_c$, a circumstance evident as, well in the Cu NQR work from Hunt *et al* [60].
12) the level of charge ordering is sufficiently strong that there is clear *z*-axis correlation, the main superspotting developing at 0.5 $c_o^*$, corresponding to a 4-layer stacking sequence.

From this and the preceding introduction we now have accumulated enough data to construct geometrical models as a function of $p$ for the charge and spin structuring adopted.

### §4. The structural setting of local pair superconductivity in the cuprates.

A series of stripe organized structures will be presented below, constructed in compliance with the foregoing experimental input and understanding. These will be provided for the key $p$ values within the HTSC phenomenology: namely (i) $p = 0.125$ or $8/8^2$, where the stripe organization is strongest, and manifestly there acts to the detriment of HTSC; (ii) $p = 0.1563$ or $10/8^2$, where $T_c$ maximizes; (iii) $p = 0.1837$ or $9/7^2$, where the condensation energy per pair, $U_c(p)$, maximizes; (iv) $p = 0.277$ or $10/6^2$, where superconductivity terminates, and (v) $p = 0.055$ or $22/20^2$, where HTSC commences.

**(i).** $p = 0.125$ or $8/8^2$. (figures 1, 2 & 3) **Charge stripes and magnetic discommensurations.**

The $8a_o$ cell in fig.1 is the one for which, as we have seen, most data exists, the stripes/DCs here being frozen in, or nearly so, and the diffraction sharp. The domains of antiferromagnetically coupled spins stand in antiphase across the DCs, which are aligned in the 45° orientations. The magnetic spotting implies a full two-atom phase slip of the 'up/down' $2a_o$ sequence after 4 such units, or, rather, a registerable 180° phase slip of $a_o$ after just 2 such units or $4a_o$. At a general $x$ coordinate in the cell, upon moving in the $y$ direction in fig.1 we then traverse two what might be termed 'partial discommensurations' within the supercell repeat of $8a_o$. At $x$ values corresponding to the DC crossing points a y-axis traverse sees the action of the pair of DCs become fused into a single 360° phase slip.

The domains are magnetically not all equivalent but fall into two sets, 'up'-centred and 'down'- centred. The true orientation of the spins in fact remains uncertain. Some suggest that the spins lie in or very close to the basal planes [61], as in $La_2CuO_4$ itself [62], whilst others favour a c-axis orientation. The real circumstance is doubtless more complex. It is hard to imagine the bounding charge stripes will not affect the anisotropy forces locally constraining the spin orientations. Probably the stripes affect also local spin magnitudes. Recall μSR experiments [7,23] indicate a 50% drop in the average moment from the $0.6\mu_B$ of the parent $p = 0$ Mott insulator and they reveal a very considerable spread in local μ values. Of course that spread is much contributed to by what is to be found inside the stripes. As drawn in fig.1, there are, per $8a_o$ cell of 64 units, 26 coordination units inside the domains, with 24 on their boundaries and the remaining 14 located inside the stripes. The holes in these stripes alternate with $d^9$ moment-quenched sites, enabling us thereby to satisfy the requirement $p = |\mathbf{q}_I'|$ for the one-eighth doping level. Note that the magnetic domain periodicity along the 45° direction is $\sqrt{2} \times 8a_o$ or 44 Å, the l.t. modulation coherence length. Note also, in particular, the face-centring character to the above discommensurate array of 45°-rotated antiferromagnetic square domains, as highlighted by the shading. This is the origin of the experimental absence of any magnetic spotting in the 45° orientations.

If the phasing *along* a charge stripe above were to be slipped so that the stripe crossing point now becomes a hole, the dopant count per $8a_o$ cell would drop from $^8/_8{}^2$ to $^6/_8{}^2$, which does not then fulfil the Yamada requirement $p\ (= n_h/A) = |\mathbf{q}_I'|$. However, if the two dopant holes shed are now returned symmetrically to the $8a_o$ cell at its domain centres, we arrive at the situation portrayed in figure 2. The structure reached is the one for which the hole population is maximally dispersed. It represents Wigner crystallization under coulombic repulsion, a possibility long ago mooted in [1h]. However, this arrangement, despite now satisfying the Yamada relation, does not meaningfully echo the phase slip behaviour of the spin array. What is more, it does not present the necessary face-centring character within the *hole charge* array to match the X-ray diffraction results.

Note in figure 1 that the crystallographic basis to the *charge* face-centring geometry of the $8a_o$ supercell is the loose cluster of 4 holes about the stripe crossing points. The orientation of the face-centring in the hole charge array is, we observe, rotated by 45° as compared with the situation for the magnetic domains, and now its forbidden spot positions are the axial ones $^1/_8\ \mathbf{a}_o{}^*$ about the Γ points.  Where evidence of diffraction spotting from the charge array is to be gained, other than in the very weak $^2/_8$ $\mathbf{a}_o{}^*$ spots - the $\mathbf{q}_I{}''$ of earlier - is in the diffuse X-ray scattering reported by Issacs *et al* [63].  The latter develops in the diagonal directions as an *x/y* sum of the forbidden $^1/_8$ $\mathbf{a}_o{}^*$ axial spots.  Their work in fact was on a sample with a *p* value of only 0.075, and at this low and 'non-special' concentration the stripe organization is going to be less well perfected, the disorder greatly smearing then the recorded diagonal diffraction.

Figure 3 provides now a convenient summary of the real space and reciprocal space situation described above, contrasting the diffraction behaviour yielded by the spin and charge aspects to the striping.

**(ii).**     *p* **= 0.1563 or $^{10}/_8{}^2$.**  (figure 4).    **The optimization of *T*$_c$.**

The array witnesses the central site in the previous loose clusters of four holes now being occupied by a further hole.  The hole number per cell thus grows by 2, but the cell size does not change, as is evident from Niemöller *et al*'s X-ray work [59], from the early work of [64], and again more recently.  Yamada *et al*'s results [53] showed that the simple *p* (= n$_h$/A) = |$\mathbf{q}_I{}'$| relation is relinquished above *p* = $^1/_8$ to be replaced by a more complex progression of favoured arrangements.

From my own point of view the superstructure portrayed in fig. 4 is highly favourable to HTSC.  It establishes five-centre clusters of holes into the heart of which the four axial Cu$_{II}$ centres are able to offer low velocity *x/y* access for spin-opposed electrons.  The Madelung potential at the central hole site has become now strongly trivalent enabling it to form an electron pair production factory after the manner outlined above and throughout [1].  The embedded set of 9 coordination units constitutes the appropriate cluster with which now to attempt to go forward to a cluster dynamic mean-field theory evaluation [65] of the pairing process proposed.  It far supersedes in this regard the proposal made rather hastily in [66] for SNS2004 when first I started to consider this possibility.

I believe it not without intimation that the particular cluster pointed to above for theoretical examination presents the form it does.  The large dotted circle drawn here through the outer oxygen atoms flanking the central coordination unit is of diameter √5$a_o$.  Quintanilla and Gyorffy in [67] emerged with just such a pair interaction range (numerically evaluated as 2.3 $a_o$) when employing a cuprate band structure in their generalization of the BCS process to address more appropriately the current mixed-valent, tight-binding situation.  What that work entailed was trying to escape from the retarded, zero-range, pair interaction potential of the standard BCS approach, and to move to a finite range, instantaneous interaction.  The √5$a_o$ radius above would witness the spin-opposed electron pair being taken into the trivalent negative-*U* centre and installed largely within the oxygen sublattice.  As was discussed earlier, this is where the holes chiefly reside in the trivalent cuprates, unlike for the corresponding nickelates.  The effect has the great benefit of relieving the very strong coulombic repulsion within a pair that would arise if the double occupancy fluctuation were to be confined to the central Cu atom of the cluster.

A cluster of the above form need not owe its existence to stripe ordering, of course, it in large measure simply being an outcome of the given level of hole concentration, although striping will certainly increase its prevalence. It is important, too, to observe that at the present $p$ value the striping does not terminate with $T_c$, but persists, at least in LSCO, to appreciably higher temperatures, as is to be seen in [64]. Clearly the striping is not a product of HTSC, but rather its facilitator. In our model negative-$U$ pair production greatly contributes to the gapping of the axial, saddle-point, fermionic states under the diagonal bosonic pairing process [1c].

The value of $p$ which currently we are focussing on forms the doping level identified as being where in HTSC systems $T_c$ universally maximizes. Note it in fact falls marginally below 0.16, the number appearing in the widely used empirical relation from [16] due to Tallon and coworkers, and established through transport data such as that to be seen in [32] and [35]. By $p = 1/6$ one definitely has passed beyond the maximum in $T_c$. By then, as the neutron data of ref [30b] reveal, we have transferred to a new favoured supercell size, $q_I'$ becoming $1/7\, a_o^*$.

(iii)    **$p = 0.1837$ or $9/7^2$.**   (figure 5)   **The optimization of $n_s(p)$, $U_c(p)$ and $\chi(p)$.**

The new arrangement is gained upon reducing the previous cell size from $8a_o$ to $7a_o$ by way of a collapse of the loose cell-centring cluster to a dense cluster of just four contiguous hole-bearing coordination units. This contracted cluster should doubtless be somewhat more amenable to treatment by CDMFT, and it still displays, note, the $\sqrt{5}a_o$ feature pointed to above. The new hole content is precisely that for which the specific heat work of ref. [17] evaluates the condensation energy of the superconductivity (per hole) to draw up to a sharp maximum. Back where $T_c$ maximized at $p = 0.156$, $U_c(p)$ is only a third of what it is to become by $p = 0.184$. The latter hole content is additionally where further analysis of the $C_v$ data indicates the pseudogap, as extrapolated in from the normal state, to have dropped to zero. This same circumstance is in evidence in the recent Zn doping work of [32a], where the superconducting dome is found to collapse to zero at exactly this $p$ value as the Zn concentration steadily is built up.

In all pure HTSC systems, $p = 0.184$ also marks, as long known, the doping level at which $\chi(T)$ becomes temperature invariant, separating a regime at low $p$ where, due to residual antiferromagnetic coupling, $\chi(T)$ falls on cooling from the regime at high $p$ in which $\chi(T)$ climbs on cooling. The latter condition is perceived to come about as the uncoupled spins remaining come to respond individually to the measuring field. As can be seen from figure 5, it is possible within the present $p = 9/7^2$ array to pair up all the domain spins there in RVB fashion. While the superconductive pairing force itself may already have peaked, pair breaking is rendered a minimum here. Indeed, as the μSR 'boomerang' effect has demonstrated [15], the total number of pairs is at this point brought to a maximum.

It finally is here too that Komiya and coworkers claim they detect a slight anomaly in the normal state d.c. resistivity, similar to the circumstance at $p = 0.125$, although considerably weaker [68]. The magic value that they suggest of $3/16$ or 0.1875 does not however accord with their data quite so well as the figure which here we are pointing to of 0.1837.

The neutron work of Gilardi *et al* [58] employing a LSCO sample with $p$ in the present range of doping clearly supports our $7a_o$ ascription for the cell size. Often it is asserted that all such trace of stripe phase organization now becomes rapidly erased if $p$ is any further advanced towards 0.2. One suspects, though, that the circumstances might simply call for experimentation with X-rays instead of neutrons. Within the more ionic LBCO and LSCO families ordering phenomena certainly remain in evidence to well beyond $p = 0.2$.

**(iv)** $p = 0.250$ or $9/6^2$ and $0.277$ or $10/6^2$. (figure 6). **The relinquishing of HTSC.**

The recent neutron work of Wakimoto *et al* on LSCO [69] finds that by these values of $p$ a $6a_o$ cell becomes adopted. Figure 6 is drawn up for $p = 1/4$ and is of the same general form as considered previously, but now contracted down to the $6a_o$ sizing and with no hole at the central site. Occupying the latter site by yet a further hole yields $p = 0.277$, the composition for which superconductivity finally is lost. By this stage the system has become so metallized that the potential, now densely packed, negative-$U$ centres have become ineffectual in pair production, the state energy having passed appreciably above $E_F$. The ionic Madelung potentials which locally had defined the binding high valence environment no longer are sufficiently unscreened to bind the state and permit the long duration, electron double-loading fluctuations.

**(v)** $p = 0.055$ or $22/20^2$. (figure 7). **The beginnings of HTSC and the 1-q – 2-q choice.**

Above we had by $p = 0.277$ compacted down the preceding type of 2-$q$ DC structuring as far as is meaningful. As noted when assessing the X-ray work of Ghazi *et al* [55] on the corresponding nickelates, that composition would support a 1-$q$ faulting, terminating the progression of favoured structures identified in §3.3. Now at the other end of the composition scale, at the threshold to the spin-glass regime, there similarly arises clear evidence for a 1-$q$ striping becoming the favoured option. The work of Wakimoto and colleagues [70] performed on $(La_{2-x}Sr_x)CuO_4$ $x = 0.050$, as well as on 40% Nd-substituted material, clearly identifies at this point an asymmetry within the $q_I'$ spotting. Instead of a cluster of four equal intensity spots, there occurs a 2+2 break-up which clearly results from a uniaxial domain content in the now orthorhombic samples. Furthermore in these circumstances the magnetic spotting transparently becomes associated with the diagonal 45° direction, in a domaining process signifying just a single phase-slip direction. The phase-slip lines adhere here still to the 45° orientation, but now within a 1-$q$ rather than a 2-$q$ framework, and there no longer arises any face-centring of the array to complicate the diffraction outcome.

With an increase of doping, as the DCs are brought closer together, quite suddenly one finds the diffraction pattern, for the LSCO family at least, to alter its character, as four-fold symmetry develops in the magnetic spot intensities about π,π. With this change there comes the rotated $x/y$ axial orientation of the $q_I'$. Likewise the *charge* spotting vectors about {2,0,0}, the $q_I''$, take on four-fold symmetry, as an inspection of fig.2b in the X-ray work from Issacs *et al* on LSCO $x = 0.075$ will show [63]. The low temperature condition is orthorhombic still, but there is now no evidence of the observed pattern representing a multi-domain situation. One way to retain a single-$q$ account of events could be for the 1D stripes to alternate in orientation within successive $CuO_2$ planes up the $c$ axis. However there is no evidence of or reason for such $c$-axis order setting in sharply at 0.055. Indeed, with the rise in metallization, the drive to structural $c$-axis order should be weakening, and, indeed, Madelung forces would prefer to have the stripes AB-staggered, not crossed. Even in the spin-glass regime the stripe ordering was not strongly coupled to the LTO structure, tilting corrugations [70]. Circumstances can feasibly be somewhat different for the YBCO$_{7-x}$ system, where the chains provide stronger structural and electronic input, but, nonetheless, most diffraction data from YBCO remains remarkably four-fold symmetric in form.

The apparently terminal composition to support a 2-$q$ behaviour is $p = {}^{22}/_{20^2}$ or 0.055 and as represented in fig.7 it involves a 20$a_o$ cell of 75 Å. A simple orthorhombic cell having $p = {}^{1}/_{20}$ would when still in the 1-$q$ state call for a cell of size $\sqrt{2}a_o$ by $20.a_o/\sqrt{2}$. The cell that Wakimoto *et al* advance in fig. 10 of [70] in order to achieve a match with their diffraction data would appear not to be correct, it lacking, it seems, a $\sqrt{2}$ factor to carry the assignment made there of 7 $b_{ortho}$ up to 10 $b_{ortho}$.

For the situation shown in figure 7, with the its hole stripes in the 45° orientations, we perceive how quite good nodal conductivity can be retained so far towards the Mott insulating condition. Also, despite the quasiparticles being strongly drawn towards localization [71,72], the conditions for pair coupling are still met with at the stripe crossing points, which transmits a vestige of HTSC behaviour right through to the spin-glass regime. The number of pairs will inevitably be small, because of the limited number of carriers and the now severe pair breaking from the spin array, but regardless of this one finds from the outset that $T_c(p)$ is a function displaying negative curvature.

We are left now at this point to address two further important matters regarding the self-organization of spin and more especially charge in the mixed-valent HTSC cuprates. (1) What is the source of the strong short-wavelength electronic modulations which have recently been uncovered in energy-resolved STM work as one approaches localization, whether via reducing $p$ (locally or globally), or raising $T$, or under an applied magnetic field? (2) Are charge stripes really the source of the incommensurate magnetic response at $q_I(\omega, T)$ witnessed experimentally?

**§5. Checker boards and spin waves: what is their reality?**

**5.1). Checker-board behaviour as competition at the saddles regarding superconductive pairing.**

We now should look at what the form and origin are of the non-dispersive, short wavelength, electronic modulations recently to have been detected in a wide range of STM probings of HTSC systems as these are carried towards Mott localization. This may be accomplished in a global way via the level of underdoping, or one can move to a system that is more ionic by some appropriate selection of the counter-ions, both cations and anions. One additionally may with STM specifically elect to examine those nano-regions which by virtue of the statistical nature of the doping happen to stand closer to localization. The tunnelling spectrum clearly indicates where and whenever the probe falls within a nano-region locally supporting superconductivity or has passed into one where the pseudogap regime holds. It is possible, alternatively, to examine the pseudogap regime within otherwise superconducting material upon the application of a magnetic field, and looking then in the vicinity of a current vortex established around a quantized flux-line. Yet a further way to reach the pseudogap regime is to warm the sample up a little above $T_c$. The latter is a difficult route for STM experimentation, but has been successfully pursued by Vershinin and colleagues [73]. The magnetic field route was pioneered by Renner *et al* [74] and more recently has been taken by Hoffman, Davis *et al* [75]. The underdoping route has been explored by Davis and colleagues too [76,77], and their work much increased the attention paid to the strongly modulated 'checker-board' spatial images recorded once outside the superconducting nanoregions. Where such images now have become pre-eminent is in Hanaguri *et al*'s recent work [78] on $(Na_xCa_{2-x})CuO_2Cl_2$ [Na-CCOC]. The latter is one of the most ionic of HTSC systems as yet investigated and it produces remarkably sharp modulation images. These persist to high tunnelling voltages (up to 100 mV), whether employing a sample bias that is positive (electron injection) or negative (stronger signal; electron extraction ≡ hole injection). Why so much interest exists in the pseudogap condition is that towards the Mott regime the state is perceived as vying with the occurrence of HTSC, particularly concerning the way it relates to the Fermi surface hot-spot regions so crucial to superconductive pairing, whether by spin fluctuations [79] or in our negative-$U$ scenario [1].

The dispersionless periodicities detected in the various STM experiments referred to above cover a fair range from just below $4a_o$ to up towards $5a_o$, $4a_o$ being frequently quoted. Because of this, the phenomenon has become known loosely as '$4a_o$ checker-boarding'. Unfortunately this usage has spawned a large and exotic series of real-space-based suggestions for its origin [80-87]. However the limited periodicity range being reported would suggest here some *k*-space related effect. Certainly the effect is electronic in nature rather than atomic (as just might possibly have arisen from, say, some more complex J-T arrangement) since no trace is left in standard X-ray or neutron crystallography. One very pertinent observation is that the more underdoped a sample is the shorter becomes the primary modulation wavelength reported and the more ionic the system is the sharper its definition. In the Na-CCOC case [78] and for samples to either side of the threshold to a delayed superconductivity around $p = 0.07$, the effect has become so strong and the images (obtained at 10 mK and positive sample bias) so complex that they readily support a full Fourier analysis of what is afoot. Such analysis shows that the images emerge in this case predominantly from an admixture of *k*-values of $^1/_4\mathbf{a}_o^*$, its complement $^3/_4\mathbf{a}_o^*$, and $\mathbf{a}_o^*$ directly [78]. The shortest of these *k*-vectors could suggest the separation between quasiparticle states across the individual arms of the Fermi surface near the saddle points – exactly the hot-spot regions. However there at first sight exists a considerable bar to such an ascription – namely the k-space span across the arms is from LDA band structure calculations [88-90] not this large, in ordinary circumstances. A simple construction based on a circular Fermi surface centred upon the zone corner would support an arm spanning vector at $p = 0.12$ of only $^1/_{6.4}\,\mathbf{a}_o^*$, not $^1/_4\,\mathbf{a}_o^*$. Recent detailing by ARPES [91] and STM [8,1a] of the Fermi surface geometry however would assert that the real situation is otherwise and accordingly would afford relief to such an impasse. How can this be?

Some years ago photoemission results indicated that the saddle point regions of the band structure were disproportionately renormalized by correlations, consequential upon their x/y axial position, their low binding energy, and the $d_{x2-y2}$ $\sigma^*$ wavefunctions specifically involved. The early

ARPES work of Gofron, Campuzano *et al* [92], of King, Shen *et al* [93], and of Ma, Onellion *et al* [94], conducted on both optimally and somewhat underdoped material, all in fact reported that, as compared with the LDA calculations, the saddle features appear much elongated along the *x* and *y* axes, and would in such material seem to be pinned some 20 meV or so below $E_F$. Moreover more recent ARPES data, of greatly improved resolution, having the ability to separate the bonding and antibonding components arising from the bilayer interaction in BSCCO, have engendered the notion of a progressive 'unzipping' of the Fermi surface back from the antinodes towards the nodes, as the level of hole doping is diminished [11,95]. That a strong scattering- and correlation-driven gapping is being steadily introduced into normal state behaviour was first in evidence from the rising Hall coefficient seen upon cooling, and in the marked effect which the ionicity of the counter-ions was able to play here [1f,96,97], an effect reversed by the application of hydrostatic pressure [98]. A result echo-ing the Hall behaviour is the driving up of the Seebeck coefficient upon cooling [10]. The latter effect, of course, is the one which lies behind the commonly employed method to assess doping content, introduced by Obertelli, Cooper and Tallon [99].

Precisely how correlations and the approach to the Mott transition initiate pseudogapping and the ultimate gapping of the Fermi surface, starting out with the antinodes and working towards the nodes, has become a matter of considerable theoretical endeavour. Khodel and coworkers [100] were the first in their 'fermion condensation' work to indicate just how the specific band structure of the HTSC cuprates lends itself to appreciable renormalization under the action of strong correlations. The renormalization takes the form not only of a pinning of the energy of the Van Hove singularity, but also the straightening out of the Fermi surface geometry at the saddles, in conjunction with an increase in the lateral extent in *k*-space of the states occupied there. This same type of deformation of the Fermi surface geometry at the saddles has been arrived at as well now through the very different approach of cluster DMFT by Civelli *et al* [101], and again by Carter and Schofield [102]. An expanded Fermi sea around the hot spots and its pathologically straightened form there are precisely what we were seeking in order to account for the observed checker-boarding activity.

Note the checkerboard signal in the STM work is marked by a diminished conductance, not enhanced conduction as would come from a classical charge density wave. The injected electrons (abstracted holes) are non-propagating for the set spanning wavevector. Not only is this the case with the $^1/_4$ $\mathbf{a}_o^*$ modulation, but also for its complement, the $^3/_4$ $\mathbf{a}_o^*$ wave [78]. The relative degree of swelling and unzipping of the Fermi surface across the progression BSCCO, LSCO, and Na-CCOC of HTSC cuprates of steadily increasing ionicity is very striking [103,104]. Higher ionicity for the cationic counter-ions forces the upper oxygen-based valence band states to greater binding energies and thereby reduces the p/d mixing within the $CuO_2$ planes. This means the correlation effects will be more marked then for any particular hole doping. Conversely, under increased covalency the saddle is able to drop away from $E_F$ with underdoping at somewhat greater *p* values and for smaller departure from the M point. The binding energy associated with the saddles largely is what tunnelling and ARPES register, and like the "large pseudogap", this energy may amount to several tenths of an eV in highly underdoped material. The above features dovetail almost by accident onto the superconducting gap for samples near optimal doping. As regards the *small energy* pseudo-gapping (principally spin) active throughout in the vicinity of $E_F$, this ultimately converts into an actual gap in LSCO for *p* below 0.055. And what by that juncture has become of the lateral extent of the saddle arm? It is ~ 0.25**a**$_o$* [103], (see figure 8 to be discussed in detail below).

As was noted in the above recent comparative ARPES study from Tanaka *et al* [103], as well as in an earlier analysis by Tohyama and Maekawa [105] of Ino *et al*'s LSCO ARPES work [106], what correlation particularly influences within the band structure is the relative magnitude of *t'*, the second-nearest-neighbour (diagonal) hopping parameter. The latter, for the given wavefunctions and crystal structure, carries reversed sign in the tight-binding Hamiltonian to *t* and *t"*. The *t'* integral is governed primarily by oxygen sublattice overlap, and hence is particularly sensitive to the degree of covalence in the system.

The doping dependent evolution of events reported in Ino *et al*'s ARPES work on LSCO [106] repays close attention now. The way in which the Fermi surface modifies its conformation there in figure 7 from *x* = 0.30 to *x* = 0.05 is most revealing. The construction of that figure assumed that the Luttinger sum rule remains satisfied throughout. Such a reasonable assumption has necessarily to be embraced because for LSCO the experimental definition of $E_F$ in the spectra below *x* = 0.12 becomes very poor in the nodal directions. Recall even the nodal directions suffer significant scattering and do not provide 'good metal' behaviour [1c,9,107]. The sharper definition of the spectral onset found in the vicinity of the antinodes arises in large measure not from any well-defined metallic behaviour there, but rather from the strongly constrained DOS, with its incipient Mott gapping above $T_c$, supplanted in part by the superconductive gapping once below $T_c$. In the negative-*U* modelling of the B-F scenario, the local pairs precipitate Cooper pairing amongst the quasiparticles far around the Fermi surface in feed-back fashion.

Remember from the μSR work that once within the superconducting state, $n_s$ comes to its peak near *p* = 0.185 [15]. With reduction in *p*, a smaller and smaller fraction of quasiparticles from what was the Fermi surface become able to be drawn into the *superconductive* condensate, causing the condensate energy per gm-atom (assessed near and below $T_c$) to decline rapidly. According to the specific heat results of Loram et al [17], $U_c(0,p)$ drops from its peak value at *p* = $p_c$ = 0.1837 very sharply, and indeed in the case of Bi-2212 the results of fig. 8 in [17b] plot out as being exponential in form. Matsuzaki *et al* [108] recently have presented their LSCO specific heat results as supporting the functional form of $U_c(0,p)$ to be only quadratic. This they then decompose as $\propto p.\Delta(p)$, where in turn $\Delta(p) = p.\Delta^o(p=0.183)$. Within my own negative-*U* interpretation the higher power of $p^4$ could at first sight look more viable, here an initial factor of $p^2$ reflecting the growth in concentration of the negative-*U* centres, with the second factor of $p^2$ representing the effective reaction rate for electron-electron pairing collisions at those centres, under the tripartite negative-*U* process. However a variation faster than power law form must be anticipated when we factor in the added effect of the pseudogap and of its collapse towards $p_c$.

This extra factor mounts rapidly with $p$ from being zero at $p = 0$ to unity near $p = 0.184$. Approaching the latter composition the pseudogap magnitude has for most $\mathbf{k}_F$ become rendered smaller than the corresponding superconducting gap.

The scale constructions which I append in figures 8 ought to help to clarify the above discussed changes which correlation is imposing on the effective Fermi surface geometry as the hole doping level is altered. Note in their LSCO ARPES work for $x = 0.15$, Ino *et al* [106] assessed the nodal crossing point to be ~ $0.4(\pi,\pi)$. Such a value declares immediately that the heavily correlated surface displays reverse curvature to what for a very considerable time was considered to hold for LSCO at optimal doping. The 'boxing up and unzippering' of the Fermi surface saddles being spoken of above would, if extended in the same manner right through to the Mott insulator at $p = 0$, position the nodal point there at $0.30$ $(\pi,\pi)$ and find the corresponding arm spanning vector raised to $\mathbf{a_o}^*/3.41$. At the other limit of $p = 0.184$, upon employing a near-circular Fermi surface centred upon the zone corner, one would emerge with an arm spanning vector reduced to ~ $\mathbf{a_o}^*/7.5$. By the latter stage the checker-boarding should have evaporated, with the entire Fermi surface action dominated now by the superconductivity. Note the above outlined pseudogap behaviour is the very antithesis of Fermi surface nesting behaviour in a more customary circumstance, and appropriately it is associated with a minimum in the STM conductance measured between atom sites related by the expanded saddle spanning wavevectors.

Having introduced the general line of argument it is now time to examine some of the emergent detail more closely. Figures 8a-f have been constructed in line with the information deriving from the observed checkerboarding wavevectors, treated as above, and from the ARPES results of Ino *et al* for LSCO referred to already [106]. When the ARPES results are looked at really closely it becomes very evident that extracting hard numbers for the precise location of the Fermi level intersection wavevectors is highly problematic, even in these 11 K spectra. Indeed, of course, because, for most of the compositions studied, 11 K puts us in the $d_{x2-y2}$ superconducting regime and with the pseudogapping well advanced, such a temperature of operation introduces its own problems. Always, too, a substantial background needs to be removed that is not structureless in the vicinity of $E_F$. As Ino *et al* have stated, the particularly poor definition of the spectral onset found in the nodal direction for $p < 0.125$ means that Luttinger's theorem has routinely to be invoked. Fortunately there is relatively very little movement of the nodal intersection point with $p$, and throughout figure 8 (apart from the case of $p = 0.24$) I have chosen to hold its location constant at $0.366$ $(\pi,\pi)$. The ARPES spectrum labelled 0.37 was in fact highlighted in figure 2c of ref.[106] for $p = 0.15$, but the neighbouring curves identifying 0.41 were selected both in figure 1c and 3c of [106] for the flanking compositions of p = 0.22 and 0.05, whereas I would prefer to stick with the lower value. Indeed figure 7 in [106] becomes largely a general schematic rather than a close rendering of the detailed evolution of the Fermi surface shape. Some of the identifications offered of the intersection points in the present key $0,\pi$–to-$\pi,\pi$ saddle spanning directions are clearly dubious and poorly defined, even when following the double differentiation technique suggested. As regards my own offerings in figure 8 of the Fermi surface evolution, I have throughout these geometrical constructions adhered to Luttinger's theorem. The pseudogapping is seen as transferring ARPES spectral weight to higher binding, but leaving the $k$-space representation for the 'diluted' coherent residue in place, just bearing reduced weighting. I have chosen not to smooth out the evaluated boundary lines between occupied and unoccupied $k$ states so as to permit the proposed evolution of the demarcation line to be plainly evident. Where the Fermi level is effectively gapped over the boxed 'vertical' stretches has been marked with a broken line, but a solid line is used where the behaviour is more standard. From the figures one sees how up to $p \sim 0.125$ the Fermi arc spreads out rather slowly from the nodal point, whilst between $p = 0.125$ and $p = 0.184$ the 'freed' section expands very rapidly, as the 'boxed' segment shrinks back towards the zone-face saddle point. This reflects what was disclosed in the specific heat data detailing the run-away growth in $U_c(0,p)$, noted above [17,108].

The sixth curve in the set is constructed for $p = 0.24$, and it marks the point where the Fermi surface necks off at the saddle. This is where the Hall coefficient goes negative [109]. It does not, however, see the end of strongly correlated behaviour. As is well known the magnetic susceptibility actually is in the process of experiencing increased temperature dependence [110], whilst the resistivity remains sizeable and, very significantly, bears a purely $T^2$ temperature dependence to more than 50 K [111]. What is more, the HTSC behaviour extends a little further yet.

**5.2). Stripes versus spin waves; the matter of dispersion.**

The persistence of strong correlation throughout the superconducting range encourages those looking towards a spin fluctuation interpretation of HTSC to persist with a RPA type treatment of the generalized susceptibility $\chi(\mathbf{q},\omega)$, suited to a more standard metal like chromium or $VSe_2$, through into the more localized regime where HTSC flourishes. I cannot believe this is a legitimate extension, and indeed, as usually conducted [79], it places no low-$p$ bound upon the rise of $T_c$. Furthermore in order to gain a match with experiment it calls for the employment of a $p$-independent Hubbard $U$ value of only ≈1 eV, and this is at a time when the correlation is rising sharply towards the Mott insulating condition. In my own negative-$U$ scenario, the optical and laser pump-probe results [1d] would indicate the negative-$U$ σ* collapse just off-setting a positive $U$ value of 3 eV to yield the local pair energy resonance with the quasiparticles. These latter values for the HTSC cuprates are appropriate to being in the BCS/BE crossover regime, where the bandwidth $W$ is ~ |$U$| at the point $T_c$ is maximized [112].

The spin fluctuation approach perceives the charge and lattice response [79] as being subsidiary in HTSC matters, whereas the negative-$U$ route regards the magnetic response to be the secondary feature. As was stated previously, the former has come about because of the dominant position of the neutron scattering results within the hierachy of HTSC phenomenology. The reluctance to embrace the central role of dynamic charge striping as being behind the incommensurate spin scattering of neutrons has been truly remarkable [113,114], and not dispelled by the synchrotron mirroring of events [59]. This course of events has revolved around the interpretation of what the so-called 'magnetic resonance peak' in the vicinity of 40 meV represents in HTSC materials with $T_c(p)^{max}$ ~ 90 K. In my own reading it is simply a monitor of spin-flip pair triplet excitation, and this accounts for its energy relation to $T_c$ being around $5½kT_c$, as appropriate for strong coupling $d$-wave behaviour. The spin-flip energy is slightly less than the longitudinal plasma resonance energy of the condensate sensed in electron energy loss spectroscopy [1a,115] and in the extra kinking that this produces near the saddle point in the band structure [116]. All three features are directly involved with the carrier pairing and evaporate near $T_c$. The stripes on the other hand, like the somewhat higher energy kinking to be found in the *nodal* band dispersion run through to higher temperatures [9,113], as befits charge/lattice effects not primarily linked with the superconductivity itself. As was noted earlier these charge effects are linked through to the similar effects obtained with the non-superconducting cuprates for p < 0.055 [70] and to the nickelates [55]. Indeed, as we have seen, the charge stripes and the associated magnetic discommensurations pick up spacing periodicities in the cuprates which are not related to fermiology but to numerology. Even where events do look to become dictated by the Fermi surface geometry in the STM-registered checker-boarding [77,78], we patently are in a regime for which standard metal physics must be inapplicable.

What are we to make then of the claim that the incommensurate neutron diffraction wavevector is energy dependent, recalling what at first sight one might expect of antiferromagnetic spin-wave dispersion, but now in an incommensurate setting governed by Fermi surface nesting [see 113]. One should first recall that this spin scattering is generated only in the limited energy range where temperature and composition put one above the spin-gap state settled into at lower temperatures and energies. If the magnetic discommensuration periodicities simply follow those of the charge self-organization stripes, as presented in §3/4, one ought not to expect the wavevectors in question to display much energy dependence: certainly not to the degree expressed in figure 4 of [113]. What I would suggest actually is being encountered here ensues from two distinct sources. First the resonant peak, as spin-flip excitation, has a dispersion from π,π that is downwards toward lower binding energies following the $d_{x^2-y^2}$ origins to the quasiparticle pairing [1a,117]. Note $5½kT_c^{max}$ in LSCO is only 18 meV, and in this single layer system a much weaker, broader feature results than in YBCO and BSCCO [1a,c]. Secondly, whether from thermal or direct excitation, the charge stripe organization suffers in conjunction with the quasiparticle self-energy to the extent that many carriers lose quantum coherence, flooding the lowest possible wavelength channels. The result is transfer of much weight from the sharp IC spotting back into diffuse AF action about π,π. It is my impression that much of the apparent modal dispersion comes from such causes, together with the added possibility that some actual localization of the hole carriers decreases the contribution to the active hole number count determining the striping IC wavevector, so making the latter shrink back on π,π. At the very least it would be good to see a five-peak analysis of the neutron diffraction data replace the customary four-peak analysis (even as adjusted by an inverse correlation function κ(ω)). The fact that the striping geometries recorded dovetail smoothly onto those of the $d^7/d^8$ nickelates, and indeed onto the cuprates in the Mott insulating condition below $p$ = 0.05, surely implies that the fermiology hand is being overplayed in this matter. Our above advocated understanding of the checker-boarding phenomenology much supports this conclusion.

## §6. Conclusions

An attempt has been made to show how a further detailed body of experimental results can be presented within the framework of the negative-$U$ boson-fermion modelling of HTSC cuprates developed previously. We have concentrated here on various lattice-based phenomena, showing how these create the detailed conditions within which the unusual electronic behaviour proceeds. The strong lattice responses arise due to proximity to the Mott transition, and also because in the cuprates we are dealing with $\sigma/\sigma^*$ rather than $\pi/\pi^*$ orbitals. The latter makes Jahn-Teller effects much stronger, securing charge segregation and striping via a strain and Coulombic route rather than by Fermi surface nesting. Where something akin to Fermi surface spanning physics does arise appears to be in the checkerboard generation coming under the action of pseudogapping, and as such is actually outside the range of normal metal physics. The hole charge striping is seen to entrain a magnetic discommensuration array in the antiferromagnetism of the inter-chain domains. Both the charge and spin arrays are in fact in most circumstances 2-$q$ and not 1-$q$ in form, and each is associated with face-centred geometry. This latter feature accounts for the observed well-known patterns of diffraction spotting, both charge and spin. The arrays are in general formed only dynamically, being primarily associated with the excess dopant charge rather than the frozen dopant ions themselves. However, relatively long-lived local environments are established from both sources. It would appear that the presence of such strong (unscreened) nanostructure lies at the heart of the creation of local pairs which drives forward the superconductivity under the boson-fermion resonance attained by means of the negative-$U$ state inversion. Two cluster geometries have been identified with which to attempt to proceed to a numerical evaluation of the proposed HTSC mechanism using the new cluster dynamic mean field theory techniques currently in development. These clusters are in line with the preliminary results from existing CPA treatments of the pairing, working with standard LDA band structures. They are ones favoured by the diagonal 2-$q$ striping.

**Acknowledgement**. The author would like to thank the Leverhulme Trust for their financial assistance in making the continuation of this work possible.

**Figure Captions**

**Figure 1.** Diagonal stripe array for $p = 0.125$. Coordination units bearing a hole charge are emphasized. Spins at the $Cu_{II}$ sites in the magnetic domains are nominally labelled 'up' (╱) and 'down' (+). The array is face-centred with regard both to its hole content and also to its 45° rotated magnetic domaining, as indicated by the shading. The charge stripes (enclosed by diagonal lines) form between the antiferromagnetically ordered spin domains, and fall where the latter exhibit $a_o$ ($\pi$–)phase slippage in the AFM. This slippage occurs simultaneously in both the $x$ and $y$ directions to generate discommensurations in the spin array follow the 45° orientation of the charge stripes. Within the latter the charge holes alternate with non-magnetic $Cu_{II}$ sites. Note the face-centred arrangement of the antiferromagnetic domains is what causes the neutron spin diffraction to produce no diagonal spotting. Here at $p = \frac{1}{8}$ is the most concentrated hole array for which the simple Yamada relation $|\mathbf{q}_I'| = p$ holds.

**Figure 2.** The figure shows what would amount to Wigner crystallization of the dopant charge at $p = \frac{1}{8}$. Hole charge has been shifted within the stripes of the previous figure to occupy the cell centre and corner sites where the stripes cross. The 'surplus' holes then have been reinserted at the domain centres. This disperses the hole charges as widely as possible. It is not what occurs in the cuprates.

**Figure 3.** Real space and reciprocal space views of the 2-$q$ charge and spin structures and resulting neutron diffraction, for the case of $p = \frac{1}{8}$ with its $8a_o$ square supercell. The charge stripe locations are indicated by the 45°-oriented, double-dashed lines, and the supercell's charge basis, containing 4 holes, is shown shaded. The $8a_o$ cell is face-centred with respect to this charge basis. The discommensurations ($\pi$-phase-slip boundaries) in the magnetic array superpose upon the charge stripes as indicated. The AF domains are 'up' and 'down' spin dominated (see fig. 1), and they form two interpenetrating face-centred, 45°-rotated arrays, as indicated by the shading. The magnetic cell side is again $8a_o$.

The charge array produces $\frac{2}{8}a_o^*$ (*i.e.* $\mathbf{q}_I''$) spotting and also diagonal spotting of wavevector length $\sqrt{2}.\frac{1}{8}a_o^*$ (see Issacs et al [63]), with $\frac{1}{8}a_o^*$ itself forbidden. The magnetic array on the other hand produces $\frac{1}{8}a_o^*$ spotting (*i.e.* $\mathbf{q}_I'$), with this time $\sqrt{2}.\frac{1}{8}a_o^*$ and $\frac{2}{8}a_o^*$ forbidden. The charge satellite spotting is to be centred around the $\Gamma$ points for the b.c.t. crystal structure, $(m2,n2)a_o^*$, whilst the magnetic spotting devolves from the basic Néel structure spotting of the $p = 0$ Mott insulator sited diagonally at $(\frac{1}{2},\frac{1}{2})a_o^*$, etc.. Point (0,1) is not equivalent to (0,2) being a Z′ point in b.c.t. geometry

**Figure 4.** The proposed circumstance for where $T_c$ becomes maximized. The structure is again an $8a_o$ cell as in figure I, but now with two extra holes per cell occupying the stripe crossing points

at the cell centre and corner. Of course this cell now is richer in holes than for the Yamada relation, having $p$ = 0.1562 or $^{10}/_8{}^2$ in the $8a_o$ supercell. The 5-hole clusters are seen as constituting very favourable pair production centres within the negative-$U$ modelling adopted, the central site of the cluster being strongly $Cu_{III}$-like in nature. The large dotted circle running through the surrounding oxygen sites of the outer hole cells within the cluster is of diameter $\sqrt{5}a_o$. Calculations to be found in [67] would support this as marking the extended range of the interaction potential function to replace the $r$ = 0 δ-function employed in the classic BCS treatment of superconductivity.

**Figure 5.** The situation proposed for the stage where the pseudogap vanishes and the super-conducting condensation energy per pair, as deduced from the specific heat work [17], becomes maximized. It corresponds to $p$ = 0.1837 or $^9/_7{}^2$. The $7a_o$ cell is reached from figure 4 by collapsing the central cluster from a loose 5-hole one to a tight 4-hole one. Again the large dotted circles are each of diameter $\sqrt{5}a_o$. The spins can here all be paired up in RVB fashion, appropriate to the fact that at this hole content most electrons become engaged in the superconductivity and the susceptibility is temperature independent with pair breaking low.

**Figure 6.** The dense array of magnetic charge stripes/discommensurations reached by $p$ = 0.25. Adding a hole at the centre site of this $6a_o$ cell yields $p$ = $^{10}/_6{}^2$ or 0.278, the point at which HTSC is terminated by the high level of metallization and a negative-$U$ state that now lies above $E_F$.

**Figure 7.** Half the $20a_o$ cell and stripe array corresponding to $p$ = $^{22}/_{20}{}^2$ or 0.0550, below which composition the system passes over from axial 2-$q$ to diagonal 1-$q$ behaviour and HTSC finally is lost. The scale is half that of the previous figures. Removing one of the cluster centring holes would give $p$ = 0.0525 and removing both of course gives $p$ = 0.0500, the Yamada composition. The level of organization of these dilute cases is probably increasingly difficult to sustain as the holes are required to be at rapidly growing separation from their corresponding substituent atoms.

**Figure 8.** Scale constructions of the Fermi surface in the repeat octant of the zone at the hole doping levels $p$ = 0.056, 0.096, 0.143, 0.157, 0.184 and 0.240. Where the surface is strongly pseudogapped in the saddle region below $p$ = 0.184 that segment is indicated by a dashed line. A solid line is used nearer the nodal direction where the carriers are freer. The nodal intersection point is based on ARPES results, whilst the 'boxed' saddle region follows the interpretation given in the text of the STM checkerboarding results. The occupied and unoccupied areas of the zone are accurately in compliance with Luttinger's theorem. The large dotted circle appearing in the $p$ = 0.184 and 0.240 plots is the F.S. appropriate to free electrons at $p$ = 0.200. The boxed segment spanning modulation wavelength covers the range from range from 4 to 5 $a_o$ seen in the STM work. The curves have been left unsmoothed to permit understanding of what is presented and register clearly the evolution of form.

$p = 0.125$ or $8/8^2$

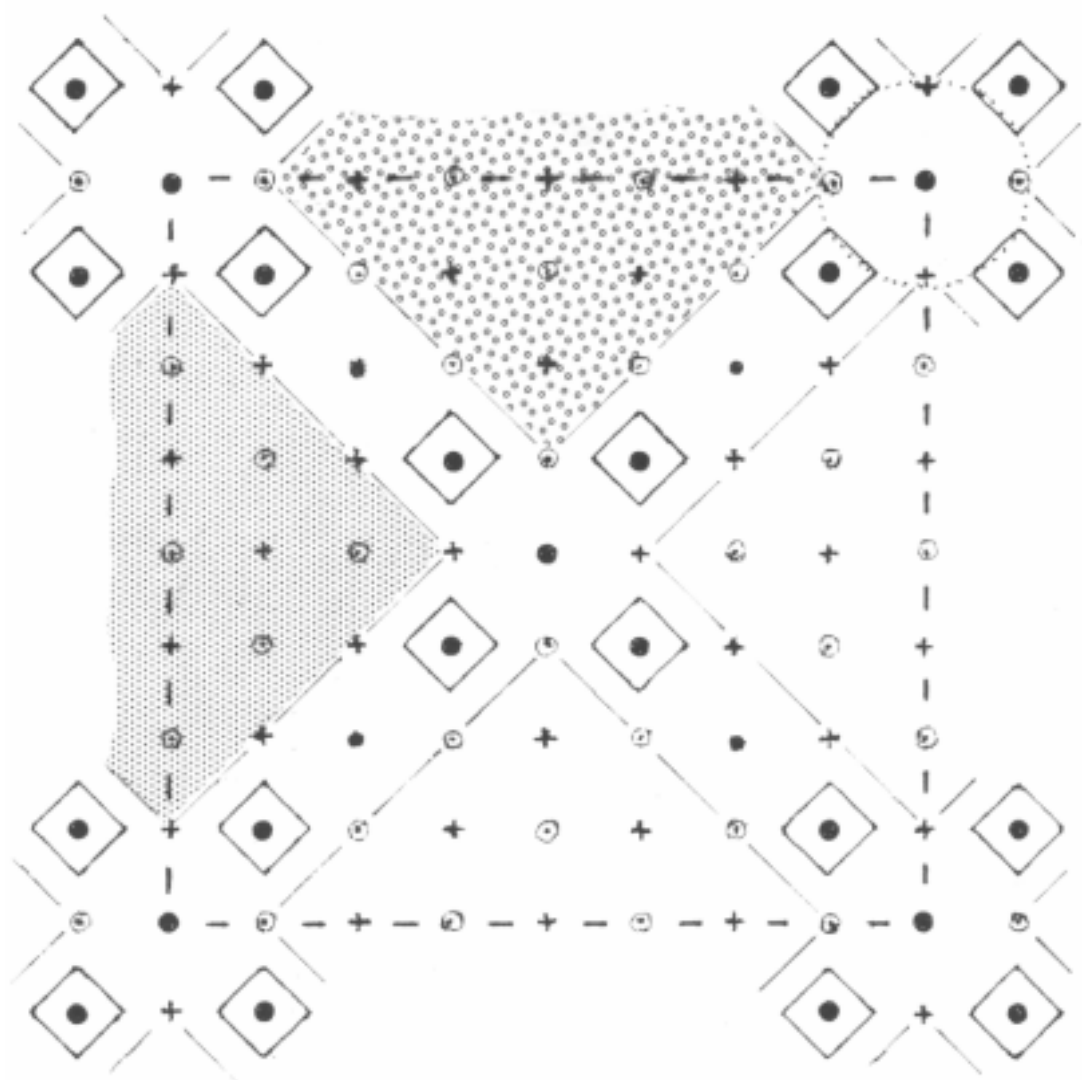

f 1

$p = 0.125$ or $8/8^2$

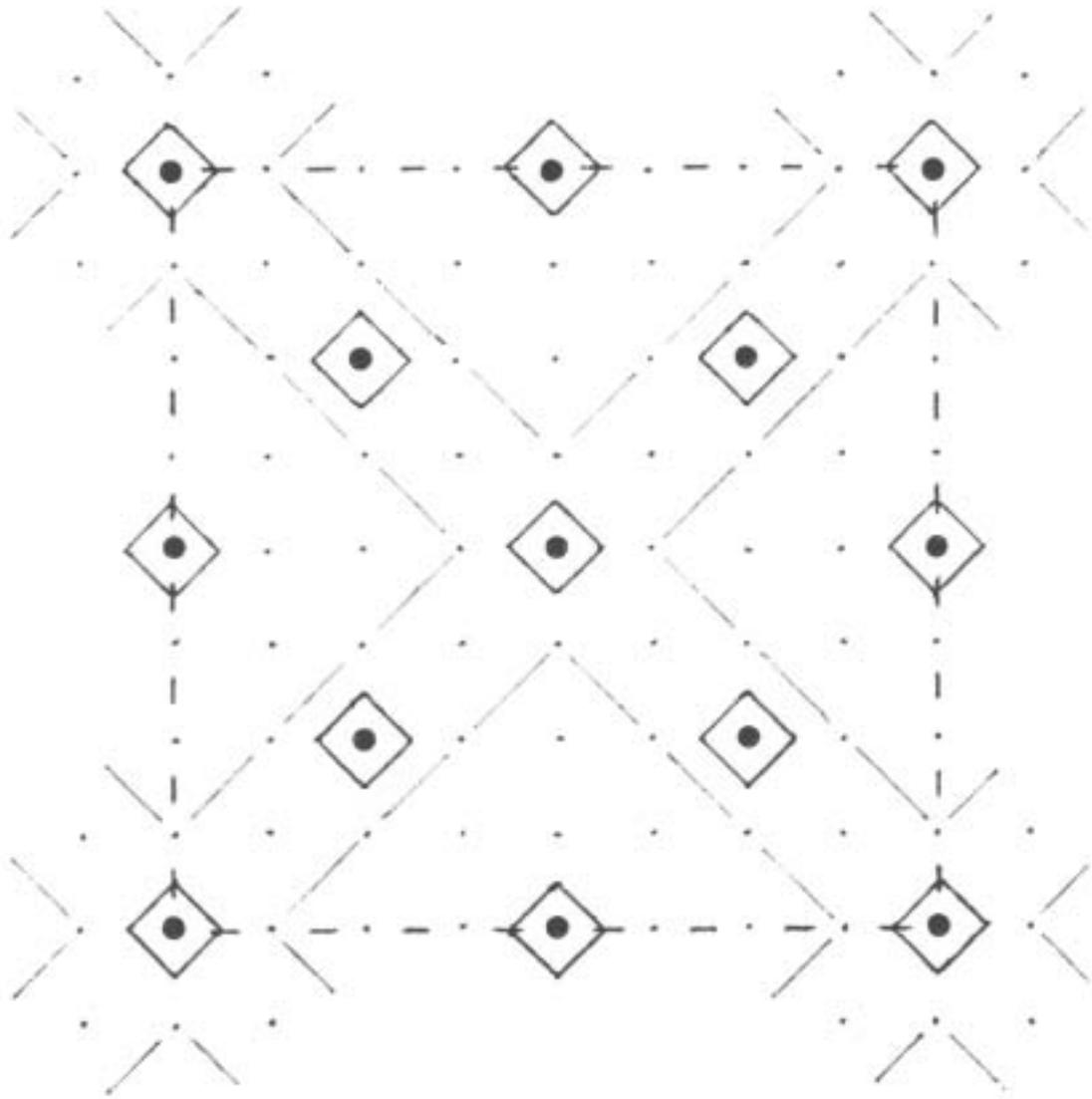

f2

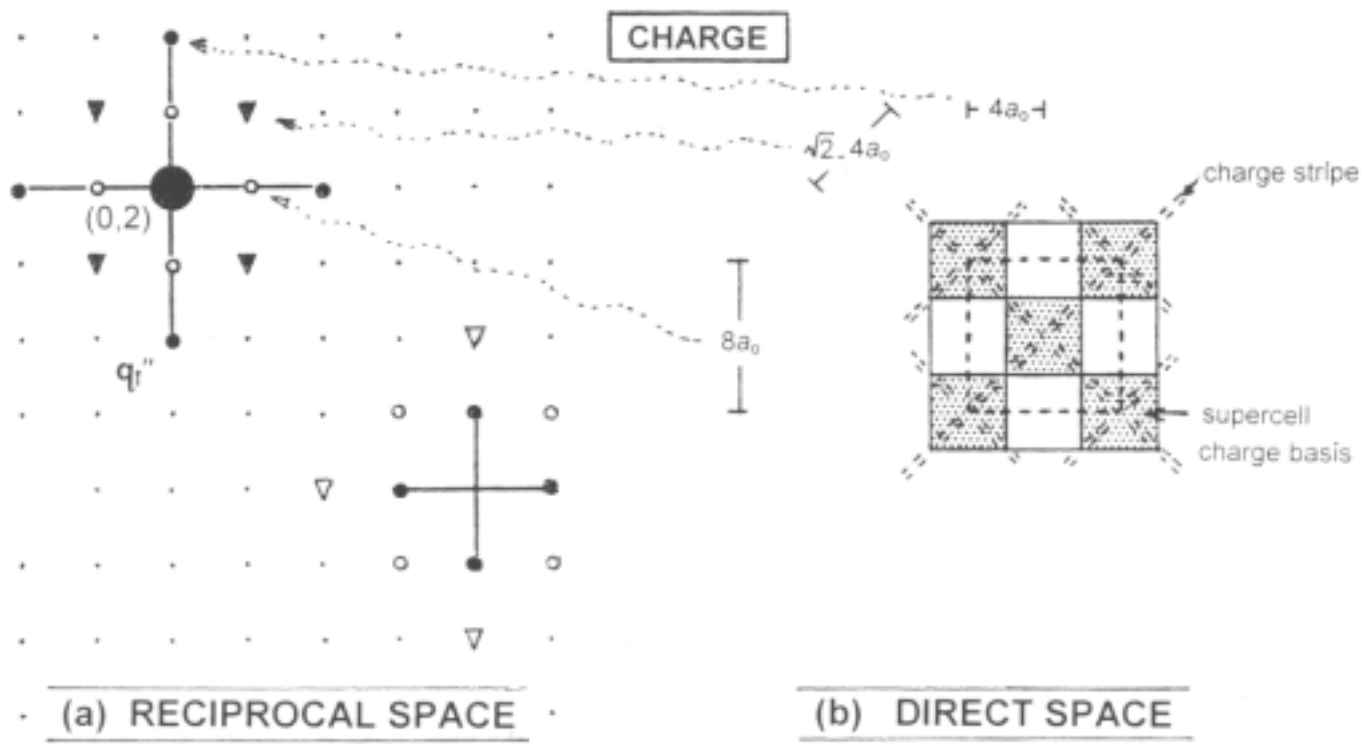

**(a) RECIPROCAL SPACE**  **(b) DIRECT SPACE**

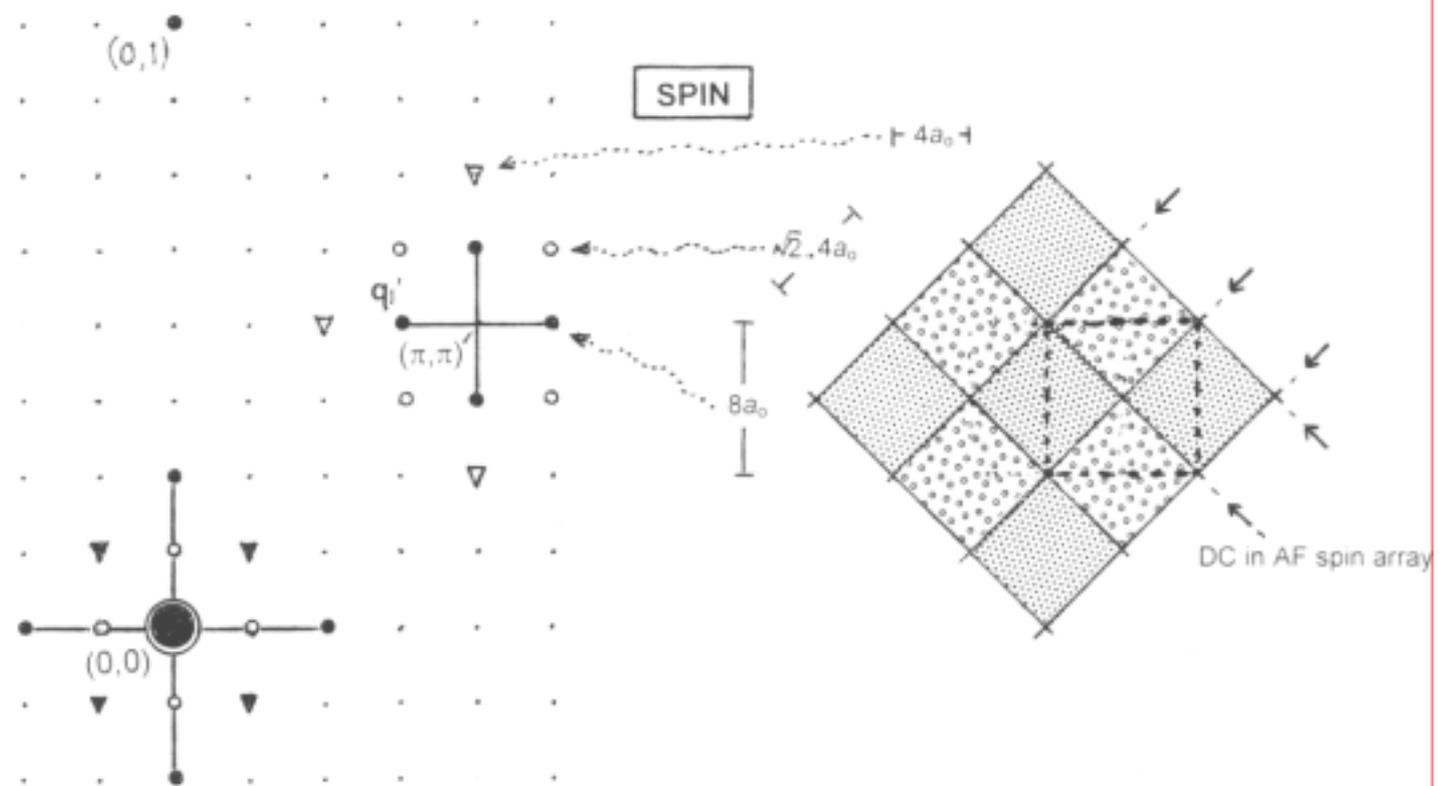

f 3

$p = 0.156$ or $10/8^2$

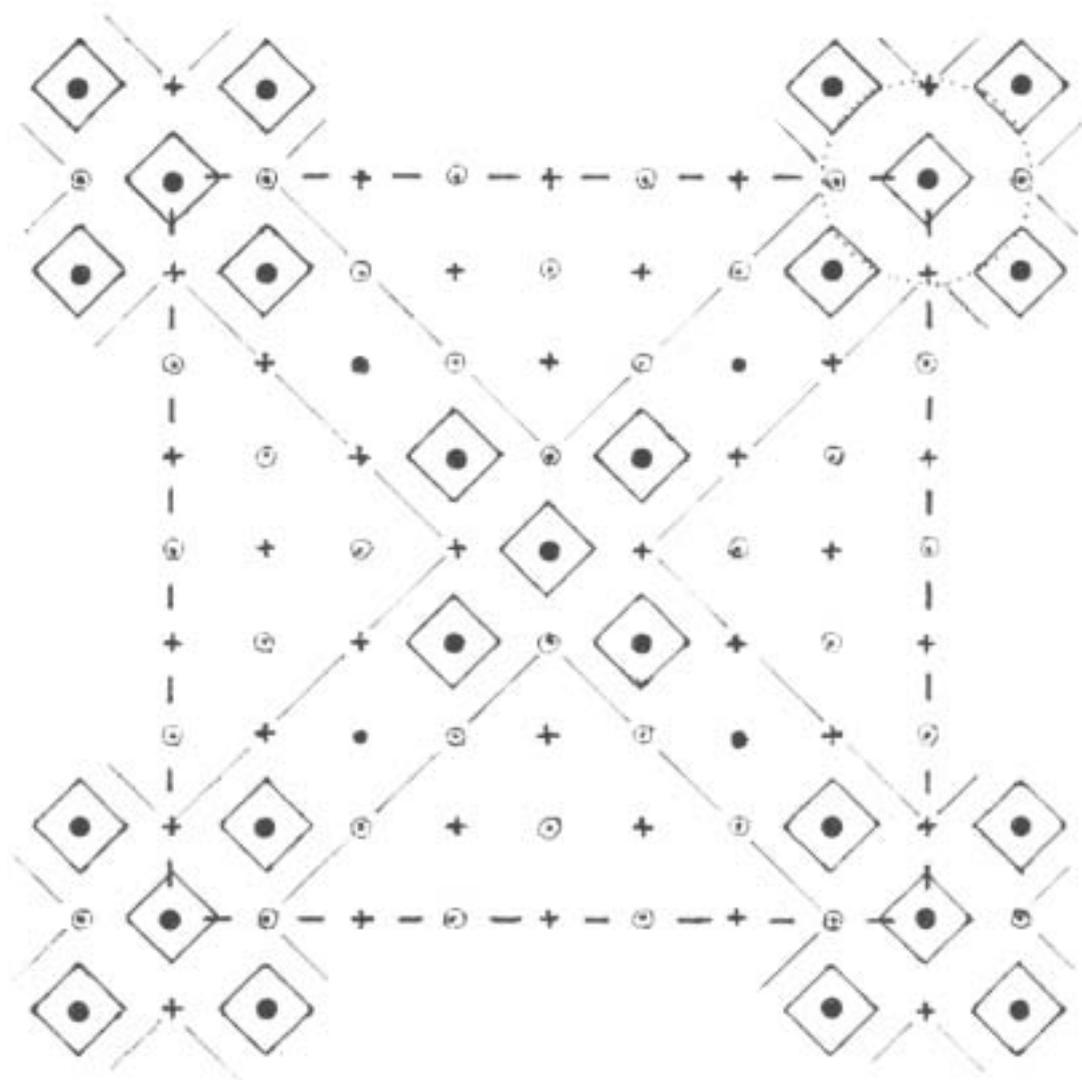

f4

$p = 0.184$ or $9/7^2$

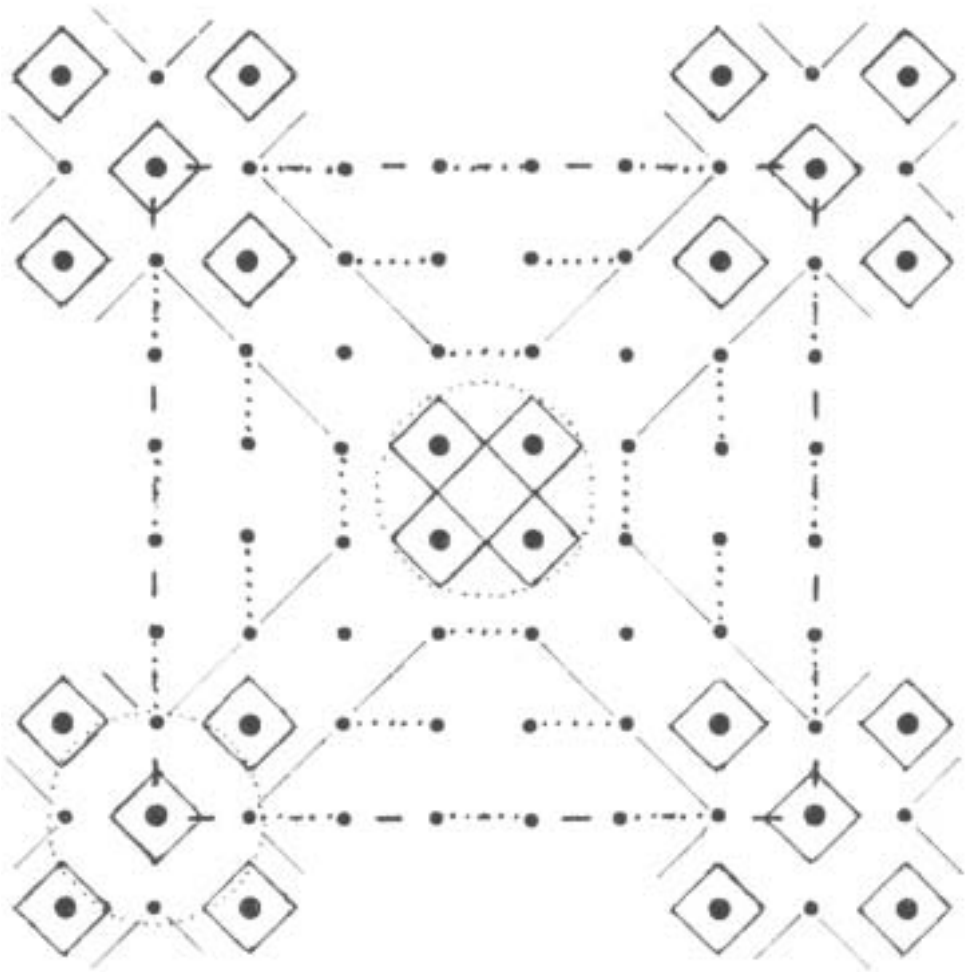

f5

$p = 0.250$ or $9/6^2$

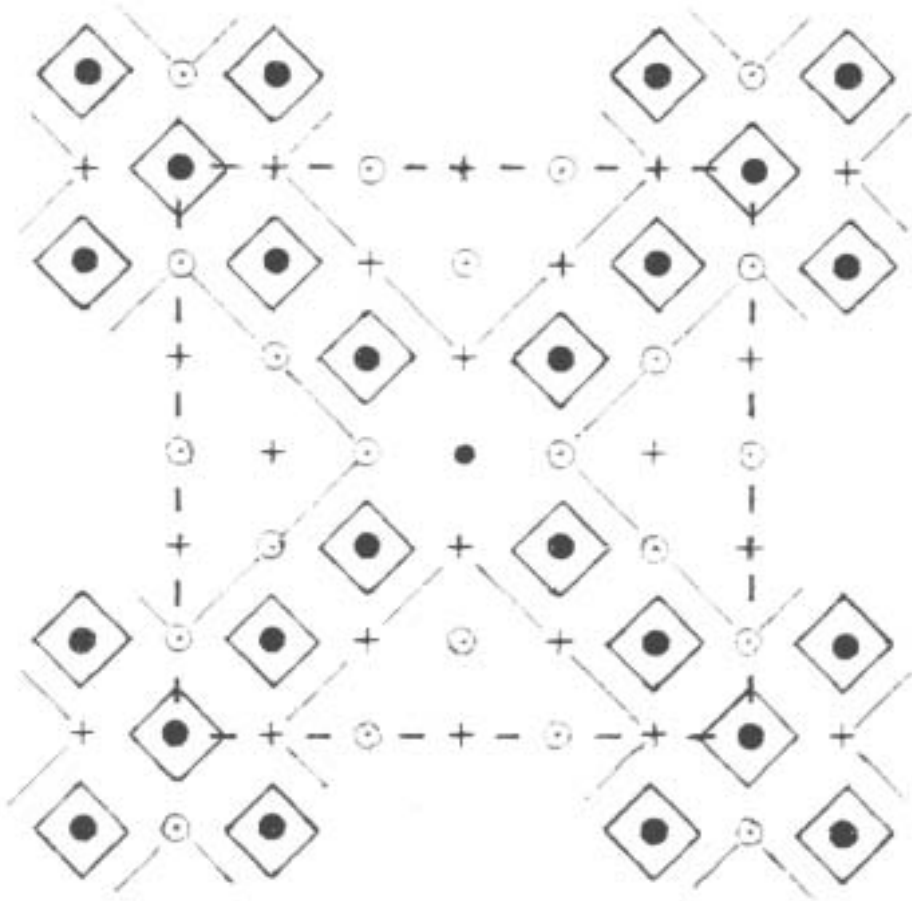

f6

$p = 0.055$ or $22/20^2$

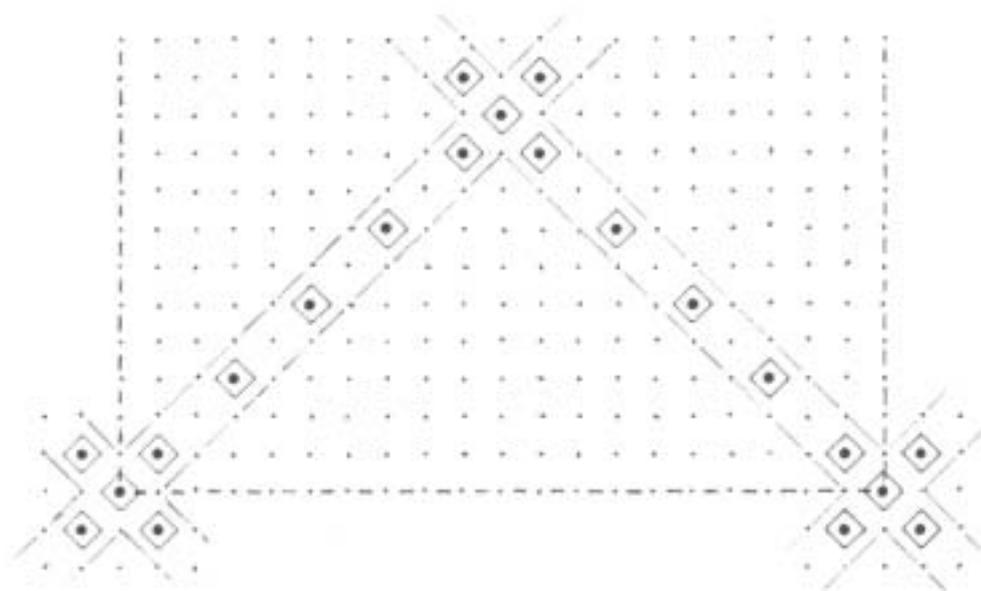

f 7

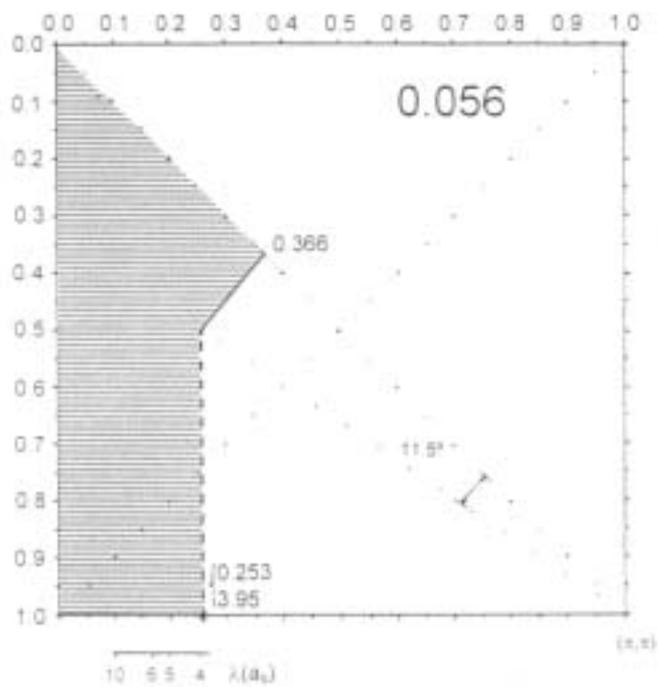
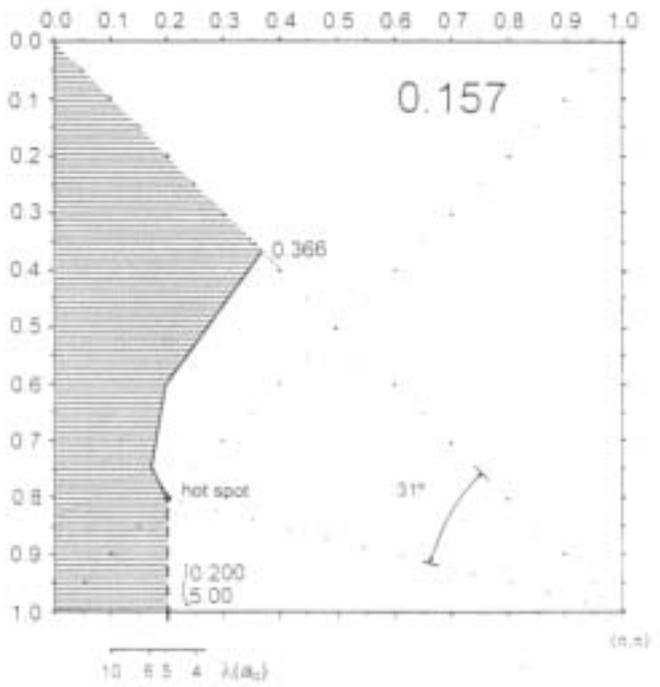
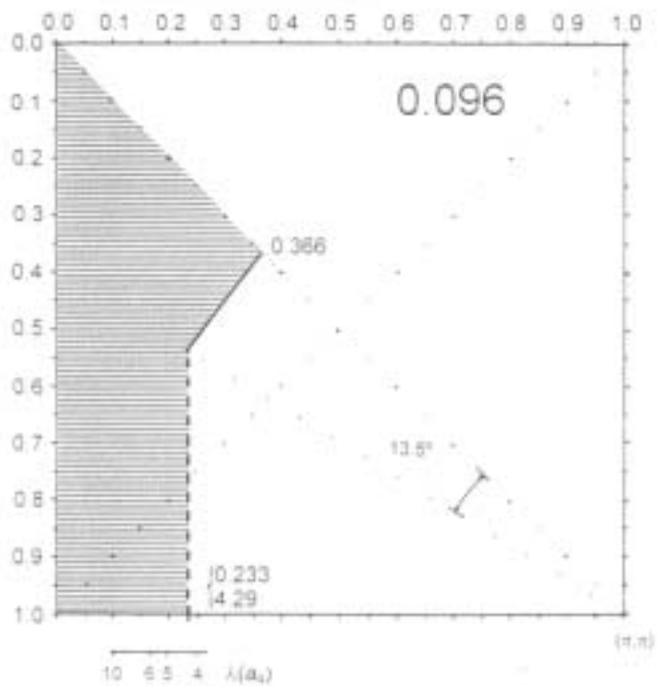
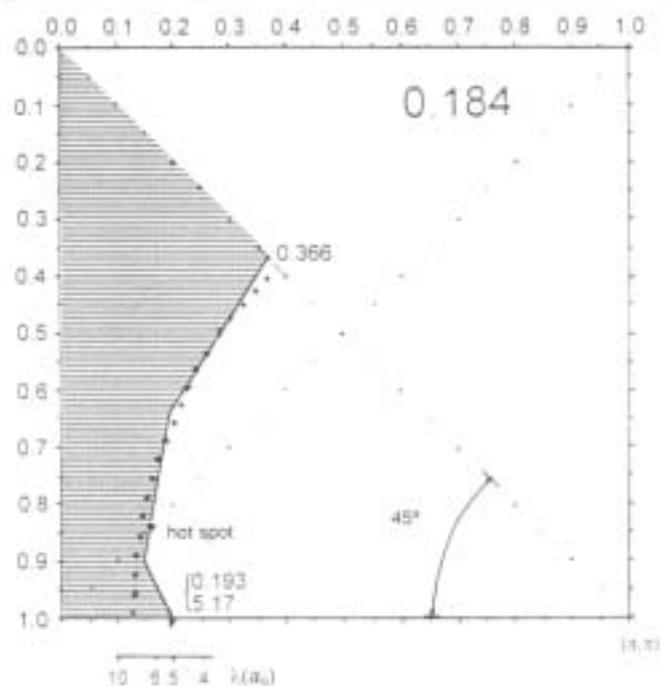
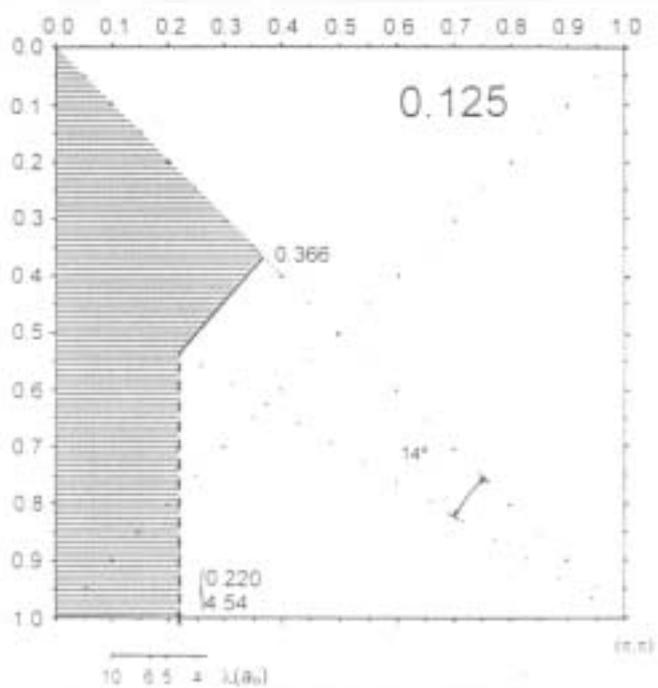
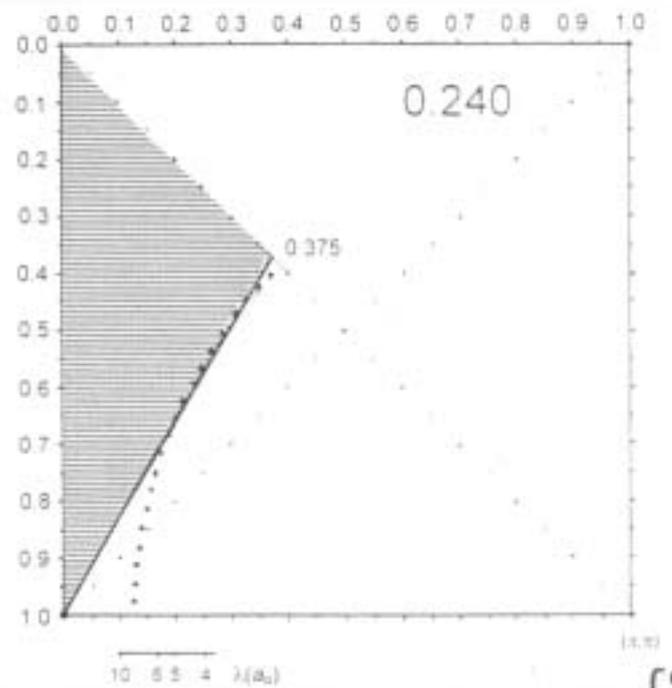

f8

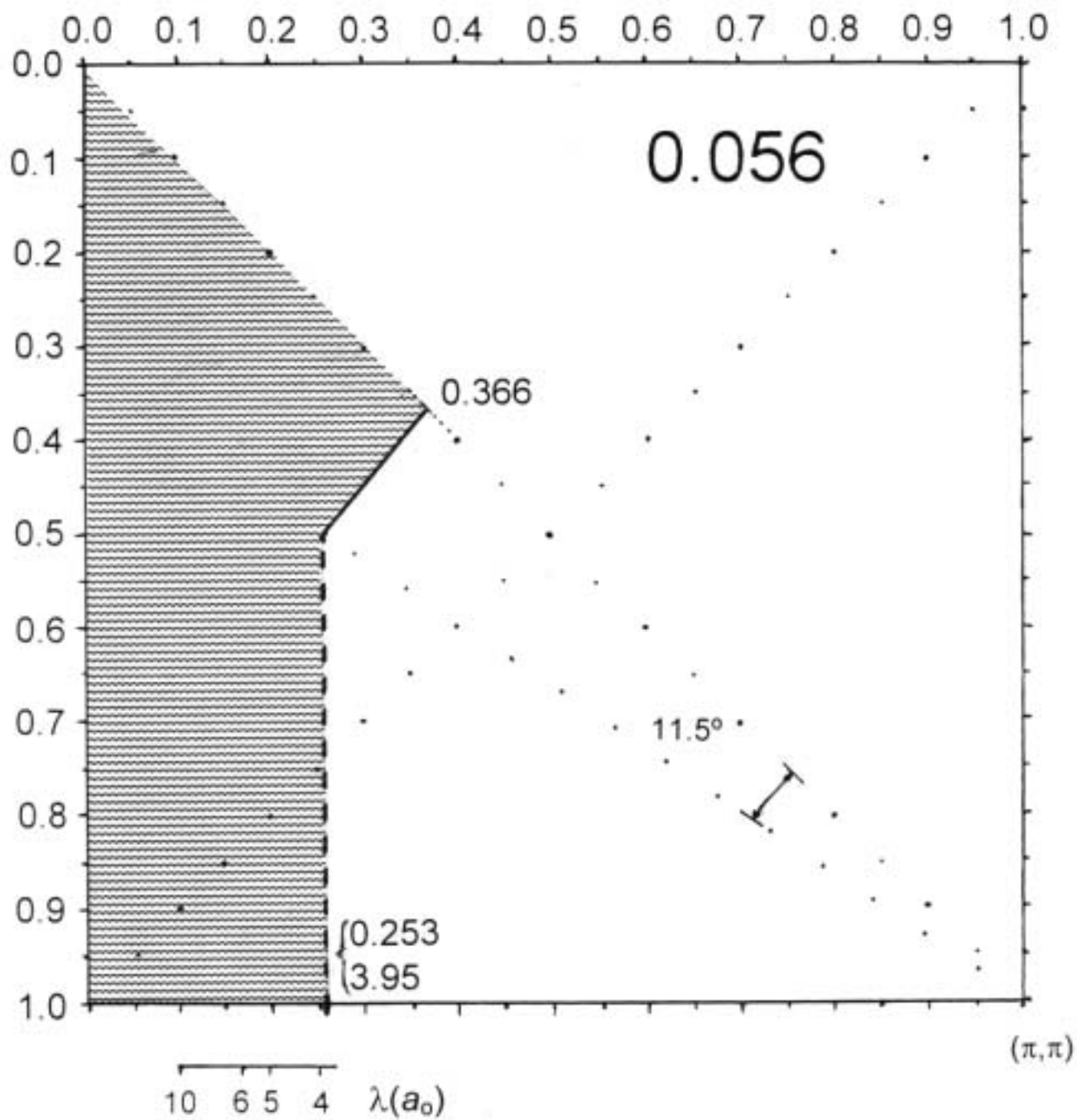

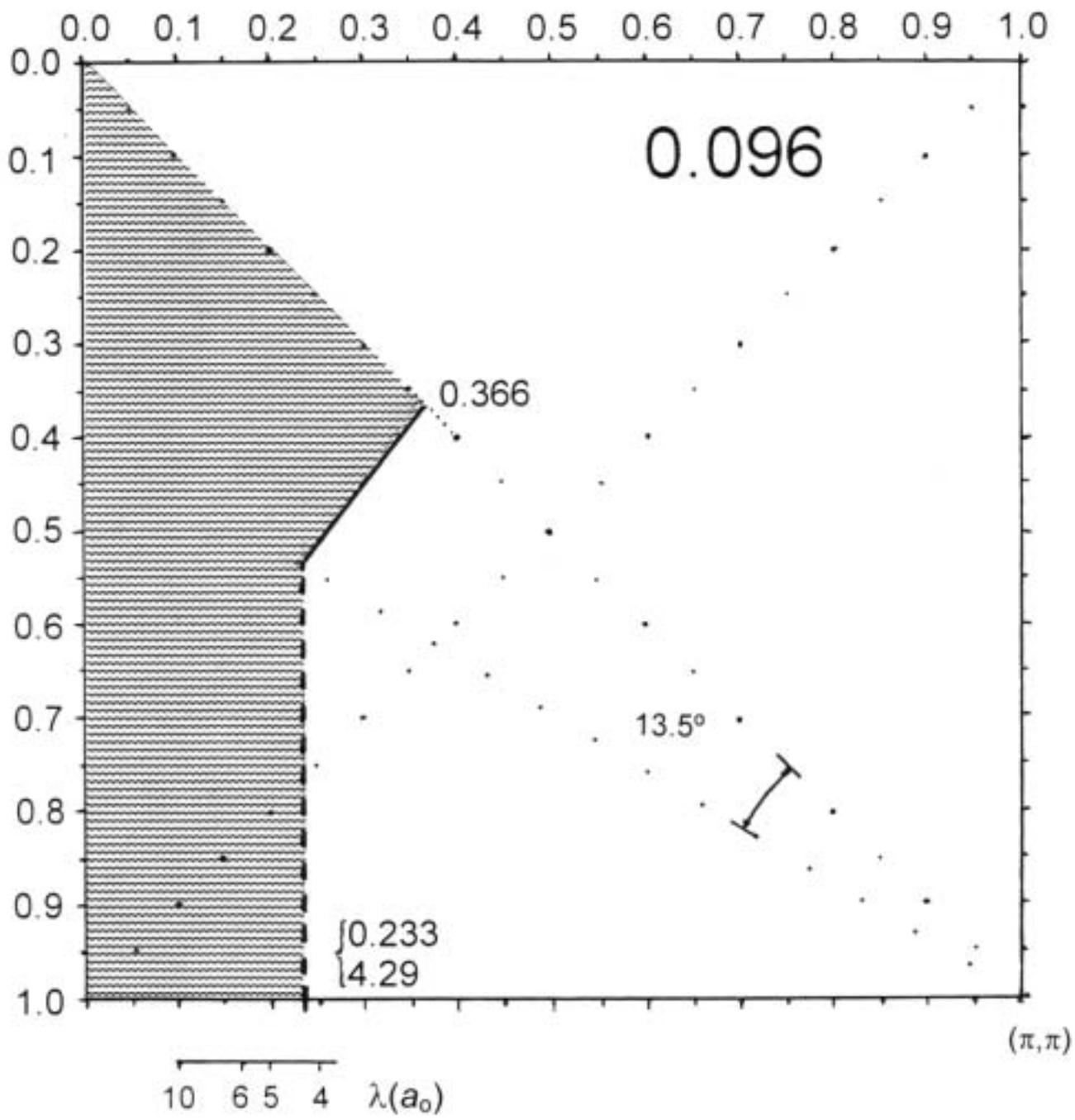

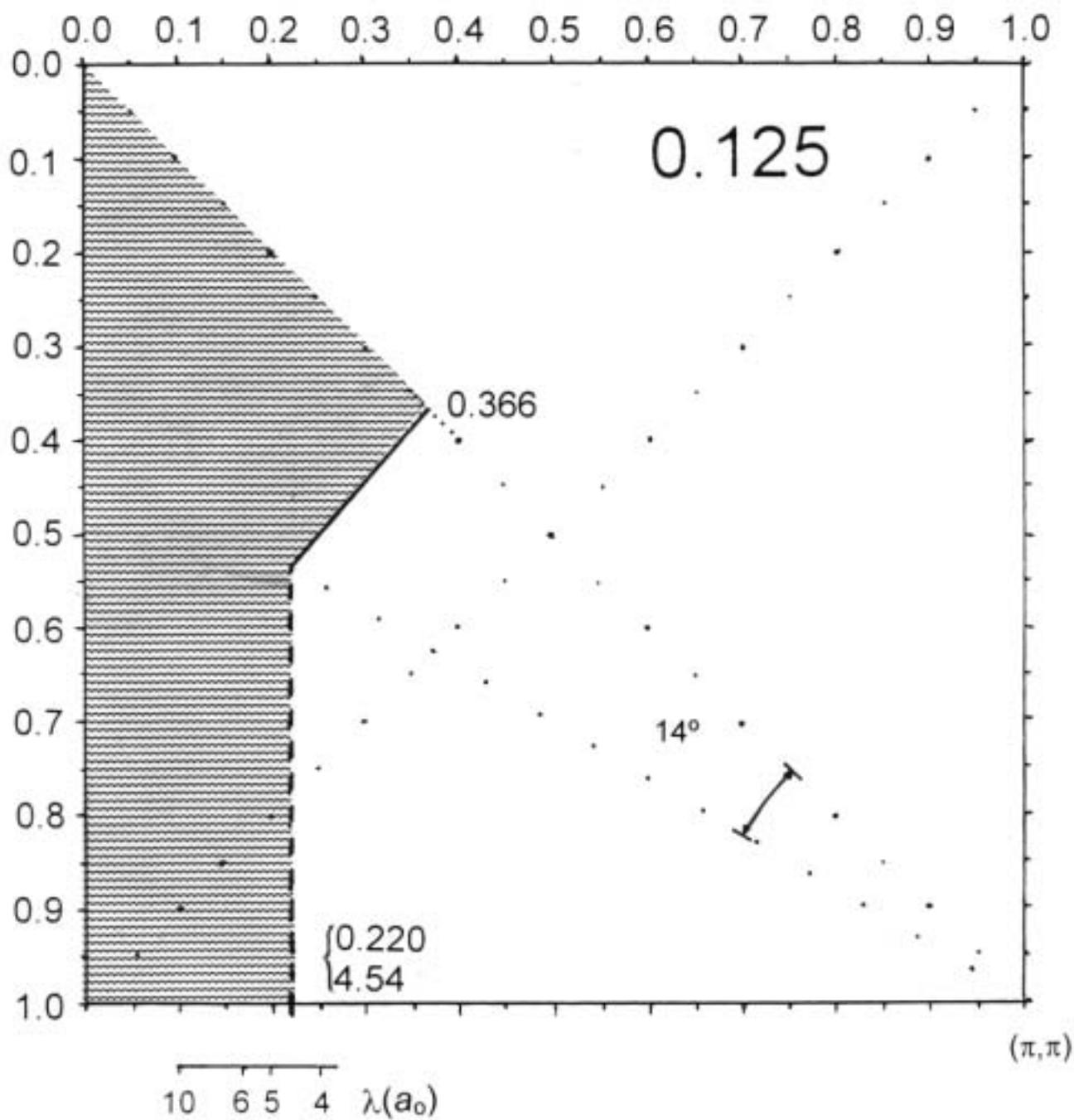

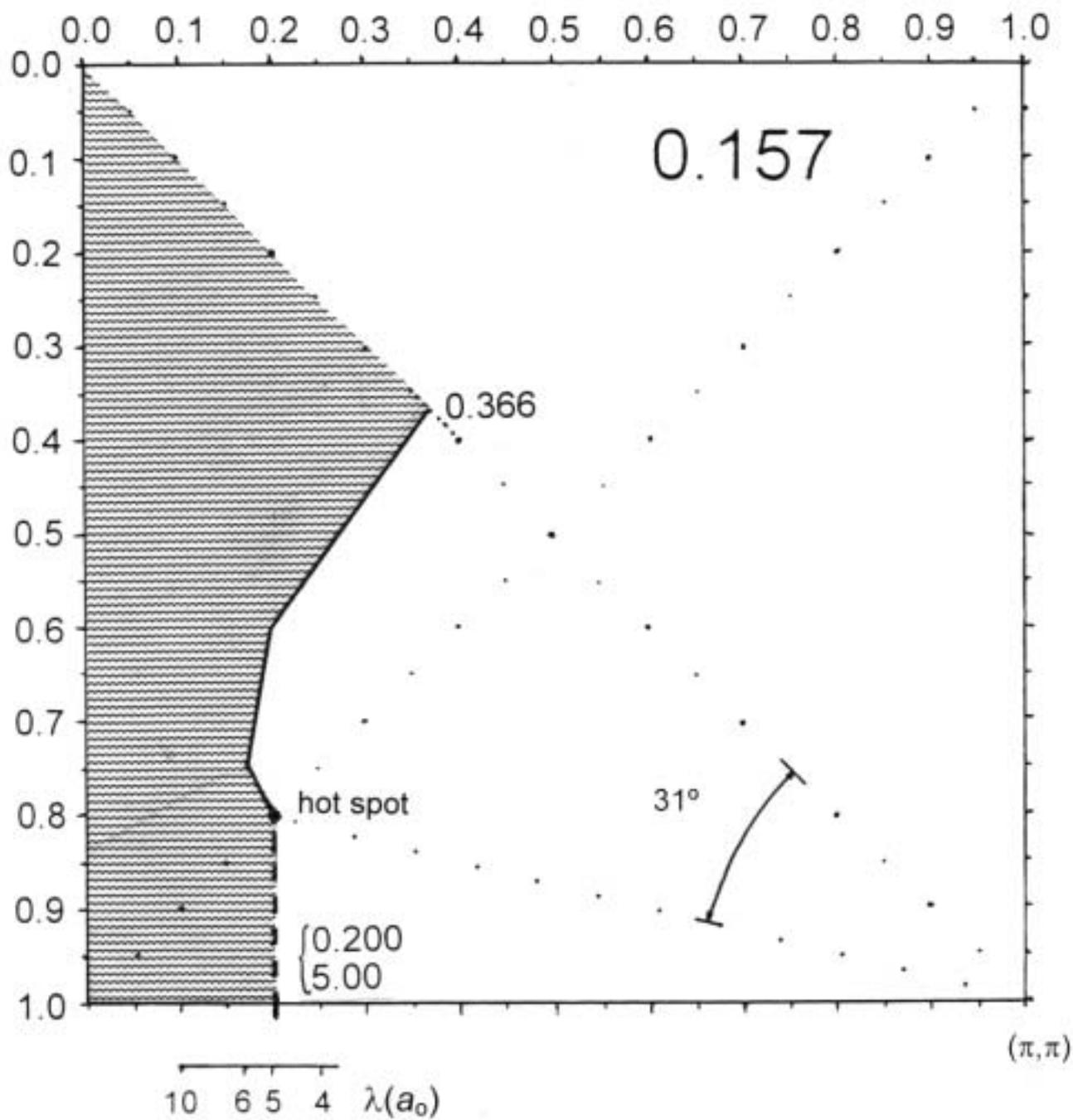

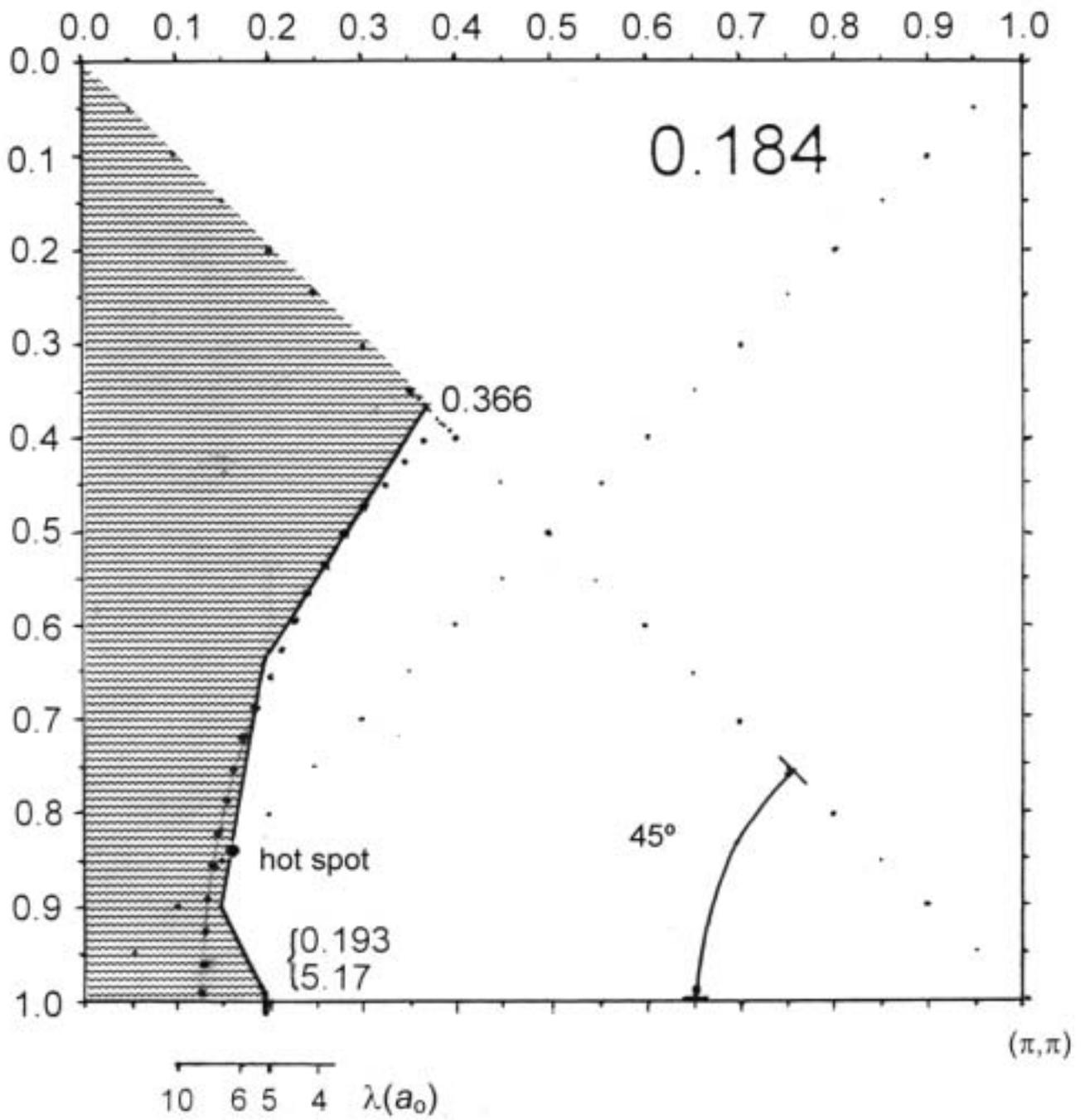

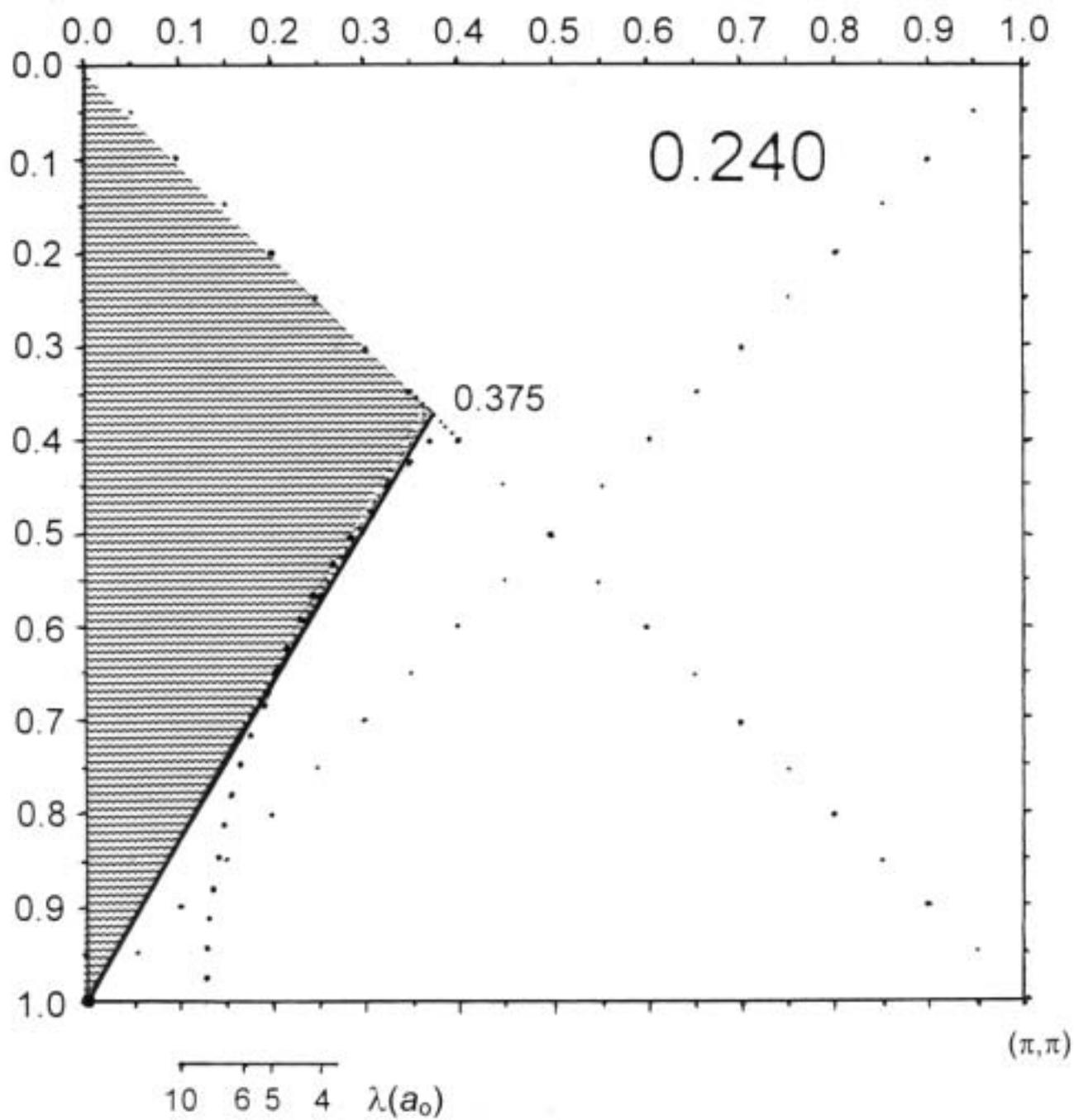

0.375

0.240

λ(a₀)  10  6 5  4

(π,π)